\documentclass[twocolumn,english]{revtex4-1}
\usepackage[T1]{fontenc}
\usepackage[utf8]{inputenc}
\usepackage{amsbsy}
\usepackage{amstext}
\usepackage{amssymb}
\usepackage{graphicx}
\usepackage{esint}

\makeatletter
 \usepackage{upgreek}

\makeatother

\usepackage{babel}
\begin{document}

\title{Universal superdiffusive modes in charged two dimensional liquids}

\author{Egor I. Kiselev}

\affiliation{Institute for Theory of Condensed Matter, Karlsruhe Institute of
Technology, 76131 Karlsruhe, Germany }
\begin{abstract}
Using a hydrodynamic approach, we show that charge diffusion in two
dimensional Coulomb interacting liquids with broken momentum conservation
is intrinsically anomalous. The charge relaxation is governed by an
overdamped, superdiffusive plasmon mode. We demonstrate that the diffusing
particles follow Lévy flight trajectories, and study the hydrodynamic
collective modes under the influence of magnetic fields. The latter
are shown to slow down the superdiffusive process. The results are
argued to be relevant to electron liquids in solids, as well as plasmas.
\end{abstract}
\maketitle

\section{Introduction}

Two dimensional electron systems are among the most studied in condensed
matter science: ultra-clean graphene sheets with impurity scattering
lengths larger than $10\,\upmu m$ \cite{Wang2013} allow the observation
of viscous electron flows \cite{deJong1995,Sulpizio2019,Gusev2020stokes,Bandurin2016,KrishnaKumar2017},
which were predicted almost 50 years ago \cite{Gurzhi1963,Gurzhi1968}.
Twisted bilayer graphene is on its way to become an important model
system for strongly correlated electrons \cite{Bistritzer2011moire,Cao2018_TBG_superconductor,Cao2020strangeTBG},
and unconventional transport effects are observed in exceedingly pure
delafossite metals \cite{Mackenzie2017,Moll2016,nandi2018_magnetotransport_delafossites,nandi2019size_effects_delafossites}.

Hydrodynamic transport theories have been successfully applied to
predict the behavior of such systems \cite{Andreev2011,Briskot2015,DAgosta2006Turbulence,delacretaz2017hydroCDW,Eguiluz1976hydrodynamicPlasmons,Forster,galitski2018dynamo,grozdanov2019holography,Holder2019,Hui2020,Link2018Out,Lucas2018,Lucas20182D,LucasH2017,Moessner2018,Narozhny2017,Principi2015,Scaffidi2017,Svintsov2018,Svintsov2019,Varnavides2020anisotropic_electron_hydro,Zdyrski2019,lucas2015memory,buchel2019_holographic_hydrodynamic}.
A particularly intriguing trait of hydrodynamics is its universality.
It can be derived from general symmetry principles without knowledge
of the underlying microscopic theory. This makes the hydrodynamic
approach particularely interesting for the study of systems where
no microscopic picture has yet been established, such as e.g. strange
metals \cite{Andreev2011,lucas2015memory,LucasH2017}.

In this paper, the mode spectrum of a charged two dimensional liquid
with weakly broken momentum conservation is investigated within a
hydrodynamic framework. We find that at large scales (or equivalently
small wavenumbers), the diffusion of charges is governed by a superdiffusive
mode, which was described by Dyakonov and Furman \cite{Dyakonov_Furman_1987charge_relaxation},
and which we interpret as an overdamped plasmon. This mode is shown
to be universal in the sense that it does not depend on any microscopic
details of the system and is determined by the rate of momentum relaxation,
the charge density and the mass density alone. We also elaborate on
the how the mode arises from the Lévy flight random walks of the individual
charged particles. Intuitively speaking, diffusion in a charged system
is faster than in an uncharged, because particles tend to spread out
more due to their mutual repulsion. Using coupled Langevin equations,
we show that the particle motion is dominated by Lévy flights and
obeys heavy-tailed Lévy stable statistics. Furthermore, we study the
influence of magentic fields on the mode spectrum and find that the
superdiffusive motion is slowed down by magnetic fields.

Anomalous diffusion has been discussed in the context of Yukawa liquids
and dusty plasmas. Numerical results implied that these systems are
superdiffusive \cite{Liu2007_Yukawa_superdiffusion}, however extensive
simulations showed that the diffusion process is ultimately governed
by Gaussian dynamics \cite{Ott2009_Diffusion}\footnote{See this Ref. for a review of literature on anomalous diffusion in dusty plasmas.}.
We reach a similar conclusion. In dusty plasmas, where the charged
particles are screened by mobile background charges and pair interactions
are well described by the Yukawa potential, ordinary diffusion prevails
(see Sec. \ref{sub:Yukawa-liquids}). However, in one component Coulomb
plasmas \cite{Ott2014_One_component_Coulomb} and certain colloidal
suspensions \cite{Warren2000_Colloidal_Suspensions}, we expect superdiffusion
as described in this paper.

\subsection{Main results}

The diffusion of charges in two dimensional systems is shown to be
intrinsically anomalous due to Coulomb interactions and is described
by the fractional differential equation
\begin{equation}
\partial_{t}\rho_{\mathcal{Q}}=2a\tau\left|\Delta\right|^{\frac{1}{2}}\rho_{\mathcal{Q}},\label{eq:Results_superdiff_eq}
\end{equation}
which is derived in Sec. \ref{sec:Superdiffusion}. Here, $\rho_{\mathcal{Q}}$
is the charge density, $\tau$ is the momentum relaxation time (see
Eq. (\ref{eq:Nav_Stokes})), $a$ is a constant depending only on
the background densities of charge and mass (see Eq. (\ref{eq:Hydro_dispersion})
and below), and $\left|\Delta\right|^{\frac{1}{2}}$ is the fractional
Laplacian \cite{kwasnicki2017_Fractional_laplace,Samko1993}. Solving
Eq. (\ref{eq:Results_superdiff_eq}) with the initial condition $\rho_{\mathcal{Q}}\left(t=0,\mathbf{r}\right)=\mathcal{Q}\delta\left(\mathbf{r}\right)$,
we find (Eq. (\ref{eq:cauchy_solution})) that the charge density
follows a broadening Cauchy distribution:
\begin{equation}
\rho_{\mathcal{Q}}\left(t,\mathbf{r}\right)=\mathcal{Q}\frac{2a\tau t}{2\pi\left(\left(2a\tau t\right)^{2}+r^{2}\right)^{3/2}}.\label{eq:Results_cauchy}
\end{equation}
Eq. (\ref{eq:Results_cauchy}) was written down by Dyakonov and Furman
in Ref. \cite{Dyakonov_Furman_1987charge_relaxation}. Deriving the
Eqs. (\ref{eq:Results_superdiff_eq}), (\ref{eq:Results_cauchy})
from hydrodynamics, we show that superdiffusion prevails in virtually
all two dimensional, Coulomb interacting systems. Moreover, the superdiffusive
behavior is universal in the sense that the only parameters that enter
Eq. (\ref{eq:Results_superdiff_eq}) are $a$ and $\tau$, whereas
it is not influenced by the nature of the microscopic interactions.
In particular, we demonstrate in Sec. \ref{sub:Dirac-liquids} that
charge relaxation in quasi-relativistic Dirac systems such as pristine
graphene, and twisted bilayer graphene is also superdiffusive. Here,
$\tau$ has to be replaced by $\tau_{c}$ – the relaxation time of
charge currents.

\begin{figure}
\centering{}\includegraphics[scale=0.4]{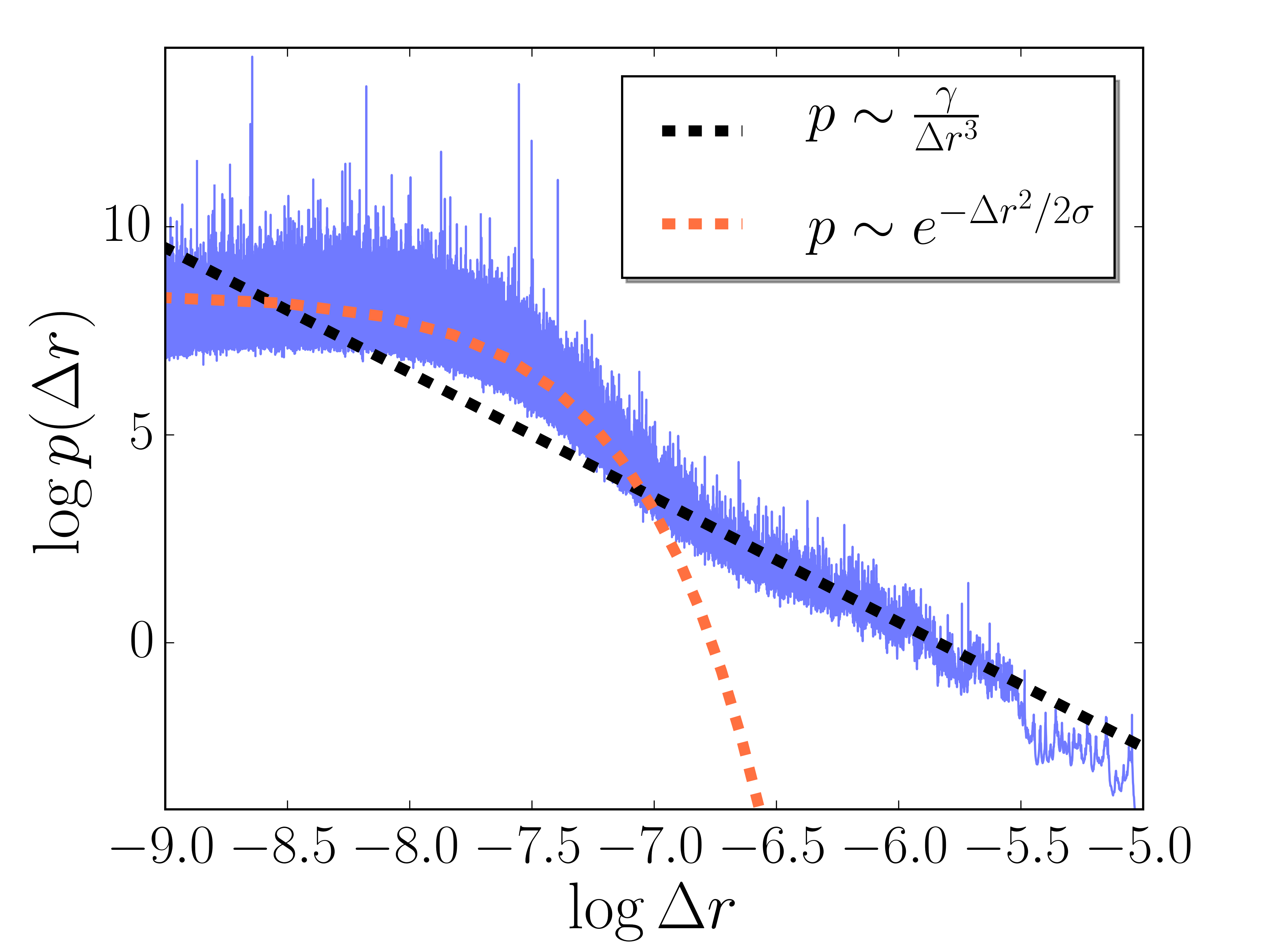}\caption{\label{fig:Langevin_x^=00007B-3=00007D_tail}The Figure shows the
step size distribution $p\left(\Delta r\right)$ of a random walk
as performed by Coulomb interacting, diffusing particles in two dimensions.
At large step sizes, the distribution clearly follows the $p\sim\Delta r^{-3}$
power-law which leads to the superdiffusive dynamics described by
Eq. (\ref{eq:Results_superdiff_eq}). The data was obtained by integrating
the system of coupled Langevin equations of Eq. (\ref{eq:Langevin_Eq}).}
\end{figure}

The Cauchy distribution of Eq. (\ref{eq:Results_cauchy}) is a member
of the family of Lévy stable distributions. The general theory of
Lévy stability (see Refs. \cite{Bouchaud1990,Gnedenko1954}) implies
that if the superdiffusive dynamics of Eq. (\ref{eq:Results_superdiff_eq})
emerges from the random motion of individual particles – a picture
that is certainly true for classical particles – the step size distribution
$p\left(\Delta r\right)$ characterizing the particles' random walks
must decay as a $\Delta r^{-3}$ power-law:
\begin{equation}
p\left(\Delta r\right)\sim\Delta r^{-3},\quad\Delta r\rightarrow\infty.\label{eq:Results_power_law}
\end{equation}
Such a slow power-law decay invalidates the central limit theorem,
so that in the limit of many random steps, the particle distribution
does not converge to a gaussian, but to the heavy-tailed Cauchy distribution
of Eq. (\ref{eq:Results_cauchy}). Such random walks are known as
Lévy flights. To demonstrate the Lévy flight nature of the charge
relaxation process, we performed a computational experiment (see Sec.
\ref{sec:Langevin-equations}). The step size distributions of diffusing
Coulomb interacting particles was studied using the coupled Langevin
equations (\ref{eq:Langevin_Eq}). As shown in Fig. \ref{fig:Langevin_x^=00007B-3=00007D_tail},
the numerical step size distribution indeed obeys the power-law (\ref{eq:Results_power_law}),
demonstrating that the particles are travelling on Lévy flight trajectories
and their dynamics is governed by Eq. (\ref{eq:Results_superdiff_eq})
\cite{Metzler1999,Metzler2000} at large scales. The distance travelled
by the particles scales as 
\begin{equation}
\mathbf{r}\left(t\right)\sim2a\tau t,\label{eq:r-t_results}
\end{equation}
which is much faster then the $\mathbf{r}\left(t\right)\sim\sqrt{2Dt}$
law of normal diffusion: the full width at half maximum of the Cauchy
distribution (\ref{eq:Results_cauchy}) broadens with a constant velocity
$v=2a\tau$. 

Studying the collective mode spectrum of the charged two dimensional
liquid, we show that the superdiffusive mode can be interpreted as
an overdamped plasmon. In the presence of momentum relaxation, the
conventional plasmon mode $\omega_{pl}=\sqrt{2aq}$ becomes purely
imaginary for small $q$ (see Fig. \ref{fig:larger_q_modes}). The
superdiffusive mode then emerges as an imaginary branch of the plasmon
dispersion relation: $\omega_{+}=-2ia\tau\left|q\right|$ (see Eqs.
(\ref{eq:gapped_decaying_dispersion}), (\ref{eq:superdiffusive_dispersion})
and (\ref{eq:Larger_q_modes})). In the context of electron hydrodynamics,
the $\sim\left|q\right|$ dependence has been predicted for a relativistic
electron hole plasma \cite{Lucas2016sound_modes_neutr_graphene} (as
e.g. realized in pristine graphene, see also \cite{kolomeisky2017relaxation}). 

We also study the collective modes in the presence of magnetic fields
(Sec. \ref{sec:Magnetic-fields}), and find that superdiffusion is
slowed down by a factor of $\left(1-\omega_{c}^{2}\tau^{2}\right)$,
where $\omega_{c}$ is the (small) cyclotron frequency (see Eq. (\ref{eq:omega_pl_mag})).
The relaxation of charges is then governed by the equation
\[
\partial_{t}\rho_{\mathcal{Q}}=2a\tau\left(1-\omega_{c}^{2}\tau^{2}\right)\left|\Delta\right|^{\frac{1}{2}}\rho_{\mathcal{Q}}.
\]
Apart from the superdiffusive mode, we derive the magnetoplasmon dispersion
at finite $\tau$. The magnetoplasmon dispersions are shown in Eqs.
(\ref{eq:omega_min_1stLim})-(\ref{eq:omega_perp_mag}) and Figs.
\ref{fig:Mag_0.2_modes}, \ref{fig:Mag_0.05_modes} (see also Ref.
\cite{Volkov2016magnetoplasmon_retardation_effects}). It is noteworthy
that the limits $\omega_{c}\rightarrow0$ and $q\rightarrow0$ are
not interchangeable and result in different dispersion relations.
This behavior is discussed below Eq. (\ref{eq:magnetoplasmon}).

In Sec. \ref{sec:Einstein_relation}, we discuss the connection between
the Einstein relation and the superdiffusive behavior. We derive the
Einstein relation
\[
D=\frac{\sigma_{\parallel}\left(\omega\rightarrow0,0\right)}{\chi{}_{\rho_{\mathcal{Q}}\rho_{\mathcal{Q}}}\left(0,\mathbf{q}\rightarrow0\right)},
\]
where $\chi{}_{\rho_{\mathcal{Q}}\rho_{\mathcal{Q}}}$ is the charge
susceptibility, from hydrodynamics, showing that the diffusion constant
$D$ appears in a diffusive pole of the nonlocal longitudinal conductivity
$\sigma_{\parallel}$. Despite the presence of the diffusive pole
$\omega_{D}=-iDq^{2}$, the equilibration of inhomogeneous charge
or current distributions is superdiffusive, and is not described by
an ordinary diffusion equation. This changes when a gate is located
in the vicinity of the two dimensional system. If the distance between
the gate and the 2D system is sufficiently small, the long range Coulomb
potential becomes subleading to the capacitance of the gate. In this
case, the relaxation of charges indeed follows a diffusion equation,
and the Einstein relation gives the diffusion constant (see Sec. \ref{sub:Gated-systems}).

Finally, the contribution of the superdiffusive mode the specific
heat of a two dimensional liquid was calculated (Sec. \ref{sec:Specific-heat}).
At low temperatures we find 
\begin{equation}
c_{V}=c_{1}T-c_{2}T^{2}+c_{3}T^{3}-\mathcal{O}\left(T^{5}\right),\label{eq:Result_specific_heat_superdiff}
\end{equation}
where all higher order terms are of odd powers in $T$. The coefficients
are given by $c_{1}=\frac{q^{*}}{24a\tau}$, $c_{2}=\frac{3\zeta(3)}{8\pi a^{2}\tau^{2}}$,
where $q^{*}$ is a momentum cut-off. The coefficent in front of the
$T^{2}$ term does not depend on $q^{*}$. This term is due to the
non-analyticity of the superdiffusive mode at $q=0$ (see Eq. (\ref{eq:non-analyt_t^2})).
The result (\ref{eq:Result_specific_heat_superdiff}) is very different
from the $\sim T^{4}$ specific heat of the undamped plasmon mode
$\omega_{pl}=\sqrt{2aq}$ \cite{Hoepfel1982_Thermal_plasmons}. For
a normal diffusive mode in two dimension we find
\begin{equation}
c_{V,g}=d_{1}T+d_{1}'T\log\left(\frac{1}{T}\right)+d_{3}T^{3}+\mathcal{O}\left(T^{5}\right).\label{eq:Results_Diffusive_SH_contr}
\end{equation}
The $T\log\left(1/T\right)$ contribution, which is dominant at low
temperatures is not uncommon for two dimensional systems. In Sr$_{3}$Ru$_{2}$O$_{7}$,
the $T\log\left(1/T\right)$ contribution has been observed experimentally
\cite{Sun2018_SpecificHeat_Sr3Ru2O7}. Other mechanisms leading to
a $T\log\left(1/T\right)$ dependence of the specific heat are quantum
critical fluctuations of overdamped bosonic modes with a dynamical
exponent $z=2$ \cite{Millis1993_Temperature_quantum_critical} and
scattering between hot Fermi pocket and cold Fermi surface electrons
in Sr$_{3}$Ru$_{2}$O$_{7}$ \cite{Mousatov2020_strange_metal_Sr3Ru2O7}.

In experiments, measurements of the superdiffusive modes could involve
pump probe setups which can monitor the relaxation of charge carriers
at very short timescales (see e.g. \cite{Mittendorff2014pump_probe_anisotropy}).
If surplus charge is induced at a given point $\mathbf{r}_{0}$, this
charge will relax as described by (\ref{eq:Results_cauchy}). The
charge density right at $\mathbf{r}_{0}$ will decay accorting to
$\rho\left(\mathbf{r}_{0},t\right)\sim1/t^{2}$. For Gaussian diffusion,
the decay at $\mathbf{r}_{0}$ scales as $\sim1/t$, which is much
slower for small $t$. The optical properties of the material, measured
by the probe signals, will follow this dynamics. Time of flight measurements
can be another way to probe the superdiffusive behavior. Such measurements
are used to measure electron drift velocities in the presence of homogeneous
electric fields \cite{houston1977time_of_flight,hopfel1986picosecond_time_of_flight}
and could be used to probe the $r\sim t$ scaling of Eq. (\ref{eq:r-t_results})
(vs. $r\sim\sqrt{t}$ in the Gaussian case) directly.

The remainder of this paper is organized as follows: in Sec. \ref{sec:Hydrodynamics}
we introduce the hydrodynamic framework that is used throughout the
paper. Sec. \ref{sec:Superdiffusion} presents a derivation of the
Eqs. (\ref{eq:Results_superdiff_eq}) and (\ref{eq:Results_cauchy}).
Yukawa liquids and two dimensional Dirac systems are discussed. Sec.
\ref{sec:Einstein_relation} deals with the Einstein relation and
with gated 2D systems. The influence of magnetic fields on the spectrum
of collective modes is investigated in Sec. \ref{sec:Magnetic-fields}.
Sec. \ref{sec:Langevin-equations} presents the numerical results
on the Langevin dynamics of charged particles. Finally, Sec. \ref{sec:Specific-heat}
deals with the contributions of collectives modes to the specific
heat.

\section{Hydrodynamics\label{sec:Hydrodynamics}}

The motion of a charged two dimensional liquid are governed by the
laws of momentum and charge conservation and the corresponding continuity
equations. Introducing the flow velocity $\mathbf{u}$, the charge
denstiy $\rho_{\mathcal{Q}}$ and mass density $\rho_{\mathcal{M}}$,
we can write the charge current as $j_{\mathcal{Q},i}=\rho_{\mathcal{Q}}u_{i}$,
and the momentum density as $g_{i}=\rho_{\mathcal{M}}u_{i}$. In the
case of a Galilean invariant system, we have $\rho_{\mathcal{M}}=m\rho$,
$\rho_{\mathcal{Q}}=e\rho$, where $\rho$ is the particle number
density and $m$, $e$ are the mass and charge of the particles that
constitute the liquid. If the Galilean invariance is broken, $\mathbf{u}$
can be introduced as a field sourcing the conserved crystal momentum,
and the densities $\rho_{\mathcal{Q}}$ and $\rho_{\mathcal{M}}$
can be defined using the memory matrix formalism \cite{lucas2015memory,Forster}
(see Appendix \ref{app:Charge-and-Mass} for details).

The hydrodynamic equations that we will use in the following are continuity
equation for the charge density $\rho_{\mathcal{Q}}$:
\begin{equation}
\partial_{t}\rho_{\mathcal{Q}}+\partial_{i}\left(\rho_{\mathcal{Q}}u_{i}\right)=0\label{eq:Charge_conti}
\end{equation}
and the Navier-Stokes equation, which is the continuity equation for
the momentum density:

\begin{eqnarray}
\partial_{t}\left(\rho_{\mathcal{M}}u_{i}\right)+\partial_{j}\Pi_{ij} & =-\frac{1}{\tau}\rho_{\mathcal{M}}u_{i} & -\rho_{\mathcal{Q}}\nabla\phi.\label{eq:Nav_Stokes}
\end{eqnarray}
Here, $\Pi_{ij}$ is the momentum current tensor and $\phi$ is the
electrostatic potential. $\tau$ is the relaxation time of the momentum
density and accounts for momentum dissipation, e.g. due to impurities.
The above equations are very similar to the equations of classical
hydrodynamics \cite{LLHydro,Batchelor2000}. However, since charged
liquids are considered, we need to take care of electrostatic forces
induced by an inhomogeneous charge density. The electrostatic potential
$\phi$ in Eq. (\ref{eq:Nav_Stokes}) therefore depends not only on
externally applied fields but also on the charge density: $\phi=\phi\left[\rho_{\mathcal{Q}}\right]$.
In general one can write
\begin{equation}
\Pi_{ij}=\rho_{\mathcal{M}}u_{j}u_{i}+p\delta_{ij}-\tau_{ij},\label{eq:stress_tensor}
\end{equation}
Here, $\rho_{\mathcal{M}}$ is the mass density, $p$ is the fluid's
pressure. The viscous stress tensor $\tau_{ij}$ can be written as
$\tau_{ij}=\eta_{ijkl}\partial_{k}u_{l}$ using the viscosity tensor
$\eta_{ijkl}$. We limit ourself to isotropic systems where the viscosous
stress tensor can be written in terms of the shear viscosity $\eta$
and bulk viscosity $\zeta$:
\begin{equation}
\tau_{ij}=\eta\partial_{j}\partial_{j}u_{i}+\zeta\partial_{i}\partial_{j}u_{j}.\label{eq:bulk_shear_decomposition}
\end{equation}
The hydrodynamic equations (\ref{eq:Charge_conti}), (\ref{eq:Nav_Stokes})
describe the macroscopic dynamics of translation invariant fluids
with or without Galilean invariance without making assumptions on
the nature of microscopic interactions. Therefore, they are useful
tools to study the dynamics of strange metals and other materials
where a microscopic theory is currently out of reach \cite{Andreev2011,lucas2015memory,delacretaz2017hydroCDW}.

\section{Superdiffusion\label{sec:Superdiffusion}}

Particles undergoing normal diffusion spread in space according to
the law $r\sim t^{1/2}$. Superdiffusion, on the other hand, is characterized
by a faster dynamics: $r\sim t^{1/\alpha}$, where $0<\alpha<2$.
If we picture diffusion as a random walk of colliding particles, its
speed crucially depends on the so called step size distribution: the
distribution of distances that particles travel between collisions.
If this distribution has a finite variance, the resulting diffusion
process will always be Gaussian for large times and will obey the
$r\sim t^{1/2}$ scaling, which is a consequence of the central limit
theorem. If, on the other hand, the step sizes are distributed according
to a power law and their variance is infinite, we enter the realm
of superdiffusion. Another important notion is that of stability.
A distribution function $p\left(r\right)$ is called stable (sometimes
Lévy-stable or $\alpha$-stable) if the sum of random variables $r=\left(1/c_{n}\right)\sum_{i}^{n}r_{i}$,
with each $r_{i}$ distributed according to $p\left(r\right)$, is
itself distributed according to $p$$\left(r\right)$. It can be shown
that $c_{N}=n^{1/\alpha}$ is the only possible choice \cite{Gnedenko1954}.
Letting $n=t/\Delta t$, where $\Delta t$ is the time between collisions,
we obtain the $r\sim t^{1/\alpha}$ scaling mentioned above, where
$\alpha=2$ again corresponds to a Gaussian distribution. In this
sense superdiffusion is a generalization of normal diffusion for heavy-tailed
step size distributions with infinite variance \cite{Bouchaud1990,Gnedenko1954}.
In this section, we describe how superdiffusive modes arise in two
dimensional materials from the hydrodynamic equations introduced in
the previous section. 

In the following it will be useful to separate the densities $\rho_{\mathcal{Q}}$,
$\rho_{\mathcal{M}}$ into a homogeneous background and a small fluctuating
term:
\begin{equation}
\rho_{\mathcal{Q}/\mathcal{M}}=\rho_{\mathcal{Q}/\mathcal{M}}^{\left(0\right)}+\rho_{\mathcal{Q}/\mathcal{M}}^{\left(1\right)}\left(t,\mathbf{r}\right).\label{eq:inhomogeneous_separation}
\end{equation}
In the absence of external fields, the electrostatic potential is
determined by the inhomogeneous part of the charge density: 
\begin{equation}
\phi\left(\mathbf{r}\right)=\frac{1}{\varepsilon}\int d^{2}x'\,\frac{\rho_{\mathcal{Q}}^{\left(1\right)}\left(t,\mathbf{r}\right)}{\left|\mathbf{r}-\mathbf{r}'\right|},\label{eq:Poisson_eq}
\end{equation}
where $\varepsilon$ is the dielectric constant of the substrate.
The potential $\phi$ in Eqs. (\ref{eq:Nav_Stokes}) and (\ref{eq:Poisson_eq})
is the hydrodynamic analogue of the self-consistent potentials of
the Landau-Silin \cite{silin1958_landau_silin_theory} and Vlasov
\cite{Vlasov1938} theories. After a Fourier transform the above equation
reads
\[
\phi\left(\mathbf{q}\right)=\frac{1}{\varepsilon}V\left(\mathbf{q}\right)\rho_{\mathcal{Q}}^{\left(1\right)}\left(\omega,\mathbf{q}\right)
\]
with $V\left(\mathbf{q}\right)=2\pi/q$. We will be interested in
the system's response to small inhomogeneities at small $\mathbf{q}$.
Let us therefore sort out the higher order terms. The pressure term
in Eq. (\ref{eq:Nav_Stokes}) can be written as $\nabla p=\left(K/\rho_{\mathcal{Q}}^{\left(0\right)}\right)\nabla\rho_{\mathcal{Q}}^{\left(1\right)}$
with the bulk modulus $K=\rho_{\mathcal{Q}}\left(\partial p/\partial\rho_{\mathcal{Q}}\right)$.
Using the continuity equation for $\rho_{\mathcal{Q}}$, we find that
the pressure term is of order $q^{2}$: $\nabla_{i}p\propto q_{i}q_{j}u_{j}$.
The viscous terms $\partial_{j}\tau_{ij}$ also are of order $q^{2}$.
On the other hand, using Eq. (\ref{eq:Charge_conti}) one finds
\begin{equation}
\rho_{\mathcal{Q}}^{\left(1\right)}\left(\omega,\mathbf{q}\right)=\frac{q_{i}}{\omega}\rho_{\mathcal{Q}}^{\left(0\right)}u_{i}\left(\omega,\mathbf{q}\right).\label{eq:Charge_conti_fourier}
\end{equation}
Therefore, linearizing Eq. (\ref{eq:Nav_Stokes}) in $u_{i}$ and
$\rho_{\mathcal{Q}/\mathcal{M}}^{\left(1\right)}$ and performing
a Fourier transform we obtain, to first order in $\mathbf{q}$, 
\begin{equation}
\left(-i\omega+\tau^{-1}\right)u_{i}=-\frac{\left(\rho_{\mathcal{Q}}^{\left(0\right)}\right)^{2}}{\rho_{\mathcal{M}}^{\left(0\right)}\varepsilon}\left(iq_{i}\right)V\left(\mathbf{q}\right)\frac{q_{j}}{\omega}u_{j}.\label{eq:Nav_stokes_fourier}
\end{equation}
Being interested in the longitudinal solutions to Eq. (\ref{eq:Nav_stokes_fourier}),
we set $\mathbf{u}\propto\mathbf{q}$. The above equation then reduces
to
\begin{equation}
\omega\left(i\omega-\tau^{-1}\right)=\frac{i\left(\rho_{\mathcal{Q}}^{\left(0\right)}\right)^{2}}{\rho_{\mathcal{M}}^{\left(0\right)}\varepsilon}q^{2}V\left(\mathbf{q}\right).\label{eq:Omega_characteristic_eq}
\end{equation}
It follows
\begin{equation}
\omega_{\pm}=-\frac{i}{2\tau}\pm\sqrt{2aq-\frac{1}{4\tau^{2}}}\label{eq:Hydro_dispersion}
\end{equation}
with $a=\pi\left(\rho_{\mathcal{Q}}^{\left(0\right)}\right)^{2}/\left(\varepsilon\rho_{\mathcal{M}}^{\left(0\right)}\right)$.
In the absence of momentum dissipation, i.e. in the limit $\tau\rightarrow\infty$,
Eq. (\ref{eq:Hydro_dispersion}) reduces to the well known 2D plasmon
dispersion $\omega=\sqrt{2aq}$ \cite{stern1967polarizability}. Eq.
(\ref{eq:Hydro_dispersion}) describes a damped out plasmon mode,
which is purely imaginary below a threshold wavevector $q^{*}=1/\left(8\tau^{2}a\right).$
This purely imaginary branch of the dispersion corresponds to a superdiffusive
mode, as we will shortly see. For Fermi liquids, Eq. (\ref{eq:Omega_characteristic_eq}),
which describes the plasmon pole in the presence of disorder, has
been derived diagrammatically in Ref. \cite{zala2001interaction}.
As shown here, it can be justified on much more general grounds. Expanding
Eq. (\ref{eq:Hydro_dispersion}) for small $q$ we find
\begin{eqnarray}
\omega_{-} & \approx & -\frac{i}{\tau}+2ia\tau\left|q\right|\label{eq:gapped_decaying_dispersion}\\
\omega_{+} & \approx & -2ia\tau\left|q\right|.\label{eq:superdiffusive_dispersion}
\end{eqnarray}
Finally, there exists a transverse mode with $\mathbf{u}\cdot\mathbf{q}=0$
which is given by 
\begin{equation}
\omega_{\perp}=-i/\tau+\mathcal{O}\left(q^{2}\right).\label{eq:Transverse_mode}
\end{equation}

The dispersion relation of Eq. (\ref{eq:superdiffusive_dispersion})
describes a superdiffusive mode for the charge density $\rho_{\mathcal{Q}}$.
In contrast to the $\left|q\right|$ dependence of Eq. (\ref{eq:superdiffusive_dispersion}),
simple diffusive modes are governed by a dispersion relation $\omega=-iDq^{2}$,
where $D$ is the diffusion constant. In space time coordinates this
translates to the well known diffusion equation $\partial_{t}\rho_{\mathcal{Q}}=D\nabla^{2}\rho_{\mathcal{Q}}$.
On the other hand, Eq. (\ref{eq:superdiffusive_dispersion}), via
Eq. (\ref{eq:Nav_stokes_fourier}), leads to a fractional diffusion
equation for the charge density:
\begin{equation}
\partial_{t}\rho_{\mathcal{Q}}=2a\tau\left|\Delta\right|^{\frac{1}{2}}\rho_{\mathcal{Q}}.\label{eq:Fractional_diffusion_equation}
\end{equation}
The fractional laplace operator $\left|\Delta\right|^{\frac{\alpha}{2}}$
is defined via it's properties under the Fourier transform: $\mathcal{F}\left[\left|\Delta\right|^{\frac{\alpha}{2}}f\right]\left(q\right)=-\left|q\right|^{\alpha}\mathcal{F}\left[f\right]\left(q\right)$
\cite{kwasnicki2017_Fractional_laplace,Samko1993}. The special case
$\alpha=1$ is used in Eq. (\ref{eq:Fractional_diffusion_equation}). 

Fractional diffusion equations \cite{Metzler2000} are used to describe
superdiffusion in systems as different as random media \cite{giona1992fractional_random_media}
and financial markets \cite{scalas2000fractional_finance}, and can
be motivated by general symmetry considerations \cite{Baggioli2020Relativistic_Diffusion_Fractional}.
The fractional diffusion equation that we arrived at can be interpreted
as the continuous time limit of a stochastic process involving Lévy
flights. To see this, let us solve Eq. (\ref{eq:Fractional_diffusion_equation}).
Taking the Fourier transform of the spatial portion of the equation
and using the initial condition $\rho_{\mathcal{Q}}\left(t_{0},\mathbf{r}\right)=\mathcal{Q}\delta\left(\mathbf{r}\right)$
where $\mathcal{Q}$ is the charge, we find 
\begin{equation}
\rho_{\mathcal{Q}}\left(\Delta t,\mathbf{q}\right)=\mathcal{Q}e^{-2ia\tau\left|q\right|\Delta t}.\label{eq:characteristic_function_coulomb_diff}
\end{equation}
Here, we have abbreviated $\Delta t=t-t_{0}$. Taking the inverse
Fourier transform one obtains
\begin{equation}
\rho_{\mathcal{Q}}\left(\Delta t,\mathbf{r}\right)=\mathcal{Q}\frac{2a\tau\Delta t}{2\pi\left(\left(2a\tau\Delta t\right)^{2}+r^{2}\right)^{3/2}}.\label{eq:cauchy_solution}
\end{equation}
This function is interpreted as the probability distribution for the
distances a particle travels in a period of time $\Delta t$ starting
at $\mathbf{r}=0$, i.e. the step size distribution of a random walk.
Its mean value vanishes by symmetry and its variance is infinite:
$\left\langle r^{2}\right\rangle =\infty$, while its width grows
linearly with $t$. The distance $r$ that a particle travels therefore
scales as 
\[
r\sim t,
\]
which is much faster than for a normal diffusion processes, where
the distance scales as $r\sim t^{1/2}$.

In general, Lévy flights in $d$ dimensions are characterized by heavy-tailed
power-law step size distributions which scale as \cite{Desbois1992_2d_Levy}
\begin{equation}
p\left(r\right)\sim r^{-\left(\alpha+d\right)}\label{eq:Levy_jump_size_distr}
\end{equation}
for large step sizes $r$ \cite{Bouchaud1990,Gnedenko1954}. The word
``flight'' is used to stress that due to its slow decay for $r\rightarrow\infty$,
$p\left(r\right)$ allows for very large steps which would be extremely
improbable for normally distributed step sizes. The exponent $\alpha$
with $0<\alpha<2$ fully characterizes the Lévy stable distribution
function \cite{Gnedenko1954} and the distance travelled by a random
walker scales according to $r\sim t^{\alpha}$, as described in the
beginning of this section. Thus Eqs. (\ref{eq:superdiffusive_dispersion})
and (\ref{eq:Fractional_diffusion_equation}) indeed describe a Lévy
flight with exponent $\alpha=1$. Another way to see that Eq. (\ref{eq:Fractional_diffusion_equation})
describes a Lévy flight is to remember that the characteristic function
of a (symmetric) Lévy stable distribution is 
\begin{equation}
\left\langle e^{-i\mathbf{q}\cdot\mathbf{r}}\right\rangle =e^{-\gamma\left|q\right|^{\alpha}},\label{eq:general_charact}
\end{equation}
where $\gamma$ characterizes the width of the distribution \cite{Bouchaud1990,Gnedenko1954}.
For $\alpha=1$ this indeed corresponds to the solution of Eq. (\ref{eq:Fractional_diffusion_equation})
in Fourier space given in Eq. (\ref{eq:characteristic_function_coulomb_diff}).

\subsection{Dirac liquids\label{sub:Dirac-liquids}}

The prime example of a Dirac liquid is graphene at the charge neutrality
point. At finite temperatures, equal numbers of particles and holes
are excited, such that the system remains charge neutral. Thus $\rho_{\mathcal{Q}}^{\left(0\right)}=0$
holds. Homogeneous electric currents consist of equal numbers of electrons
and holes. However at finite wavevectors, $\rho_{\mathcal{Q}}^{\left(1\right)}=\frac{q_{i}}{\omega}j_{\mathcal{Q},i}$
holds, such that the self-consistent potential (\ref{eq:Poisson_eq})
must be included \cite{Kiselev2020}. In charge neutral graphene,
electric currents are relaxed by interaction effects since they are
not protected by momentum conservation \cite{Fritz2008}. The corresponding
relaxation time $\tau_{c}$ damps out the plasmon mode just as $\tau$
does in Eq. (\ref{eq:Hydro_dispersion}). The collective mode structure
of this system was studied in Ref. \cite{Kiselev2020}. The damped
plasmon mode is given by
\begin{equation}
\omega_{\pm}=-\frac{i}{2\tau_{c}}\pm\sqrt{\frac{vq}{\tau_{V}}-\frac{1}{4\tau_{c}^{2}}}.
\end{equation}
Here, $v$ is the electron group velocity and $\tau_{V}=\frac{2\pi k_{B}T\hbar}{\alpha N\log\left(2\right)}$
with the fine structure constant $\alpha=\frac{e^{2}}{\varepsilon v\hbar}$.
$\tau_{V}$ characterizes the strength of the electrostatic repulsion.
For small $q$ we find a superdiffusive mode
\begin{equation}
\omega_{+}=-i\frac{v\tau_{c}}{\tau_{V}}q.\label{eq:Superdiffusive_mode_Graphene}
\end{equation}
Similar physics will prevail in other charge neutral systems such
as twisted bilayer graphene (TBG), since at small wavenumbers the
electric current will always follow the dynamics $\left(\partial_{t}+\tau\right)j_{\mathcal{Q}}^{\left(1\right)}\left(t,\mathbf{r}\right)=-\nabla\int d^{2}x'\,\frac{\rho_{\mathcal{Q}}^{\left(1\right)}\left(t,\mathbf{r}'\right)}{\varepsilon\left|\mathbf{r}-\mathbf{r}'\right|}$,
which, together with the continuity equation will result in a superdiffusive
mode. The conclusion that charge relaxation in pristine graphene is
a Cauchy process was first reached by Kolomeisky and Straley in Ref.
\cite{kolomeisky2017relaxation} extending the original arguments
of Ref. \cite{Dyakonov_Furman_1987charge_relaxation}. Interestingly,
the phase space behavior of Dirac liquids is also superdiffusive \cite{Kiselev2019b}. 

An estimation of the value $q^{*}=\tau_{V}/\left(4\tau_{c}^{2}v\right)$
below which superdiffusion prevails in charge neutral graphene can
be made with the scattering times calculated in Ref. \cite{Kiselev2020}:
$\tau_{V}=2\pi^{2}\hbar/\left(4\alpha k_{B}T\ln\left(2\right)\right)$,
$\tau_{c}=\ln\left(2\right)\hbar/\left(0.8\cdot\alpha^{2}k_{B}T\right)$
and $v\approx10^{6}m/s$. Here $\alpha$ is the fine structure constant.
We obtain $q^{*}\approx3\cdot10^{5}\frac{1}{\mathrm{m}}$, where we
used a temperature of $T=50\,\mathrm{K}$ and a substrate dielectric
constant of $\varepsilon=6$, gaving $\alpha\approx0.1$. The above
value of $q^{*}$ corresponds to sample lengths of tens of micrometers.
Notice however, that $q^{*}\sim\alpha^{3}$. Thus the typical lengthscales
of the superdiffusive regime will strongly decrease for larger values
of $\alpha$, as they are typical for TBG. 

It should be noted, that interlayer coupling in Moiré systems can
lead to a non-trivial dependence of the dielectric function on $q$,
leading to unconventional screened interactions \cite{Pizarro2019TBG_screening,Goodwin2019TBG_screening_attractive}.
Such effects are not captured by our hydrodynamic theory, but could
be an interesting topic for further study.

\subsection{Yukawa liquids\label{sub:Yukawa-liquids}}

In a Yukawa liquid charges interact with a Yukawa pair-potential
\begin{equation}
V_{Y}\left(\mathbf{q}\right)=\frac{2\pi}{q+\kappa},\label{eq:Yukawa_potential}
\end{equation}
where $\kappa$ is the inverse screening length. Such a screened interaction
potential arises, when the considered charges are screened by mobile
background charges, as for example in dusty plasmas \cite{hamaguchi1994_thermodynamics_Yukawa}.
Two dimensional Yukawa liquids are widely studied (see e.g. \cite{Feng2013_Dusty_plasma_visc,kalman2004_2D_Yukawa,Ott2014_One_component_Coulomb}).
In particular, superdiffusion has been discussed \cite{Liu2007_Yukawa_superdiffusion},
but ultimately ruled out in favor of normal diffusion \cite{Ott2009_Diffusion}. 

Within our hydrodynamic model, it is readily shown that diffusion
in a 2D Yukawa liquid is indeed Gaussian. To this end we replace $V\left(\mathbf{q}\right)$
in Eq. (\ref{eq:Omega_characteristic_eq}) by the Yukawa potential
(\ref{eq:Yukawa_potential}). Instead of Eq. (\ref{eq:Hydro_dispersion})
we then obtain
\begin{eqnarray}
\omega_{+}^{Y} & = & -\frac{i}{2\tau}+\sqrt{\frac{2aq^{2}}{q+\kappa}-\frac{1}{4\tau^{2}}}.\nonumber \\
\omega_{-}^{Y} & = & -\frac{i}{2\tau}-\sqrt{\frac{2aq^{2}}{q+\kappa}-\frac{1}{4\tau^{2}}}\label{eq:Yukawa_dispersion}
\end{eqnarray}
For small $q$, Eq. (\ref{eq:Yukawa_dispersion}) reduces to
\begin{eqnarray}
\omega_{-}^{Y} & \approx & -\frac{i}{\tau}+i\frac{2a\tau}{\kappa}q^{2}\nonumber \\
\omega_{+}^{Y} & \approx & -i\frac{2a\tau}{\kappa}q^{2}.\label{eq:Yukawa_small_q_dispersion}
\end{eqnarray}
The mode $\omega_{+}$ describes normal diffusion where the diffusion
constant is given by $D_{Y}=\frac{2a\tau}{\kappa}$. If $\kappa$
is large, the bulk and shear viscosities, which also give a contribution
of order $\mathcal{O}\left(q^{2}\right)$ will enter the expression
for the diffusion constant. Eq. (\ref{eq:Yukawa_small_q_dispersion})
is therefore a good approximation, if the screening is weak, i.e.
$\kappa\rightarrow0$. In this case, the dynamics of a weakly inhomogeneous
charge distribution is described by the diffusion equation
\begin{equation}
\partial_{t}\rho_{\mathcal{Q}}=D\nabla^{2}\rho_{\mathcal{Q}}.\label{eq:Yukawa_diffusion_eq}
\end{equation}

\section{Behavior at larger wavenumbers, Einstein relation, gating\label{sec:Einstein_relation}}

\subsection{Behavior at $\mathcal{O}\left(q^{2}\right)$}

\begin{figure}
\centering{}\includegraphics[scale=0.3]{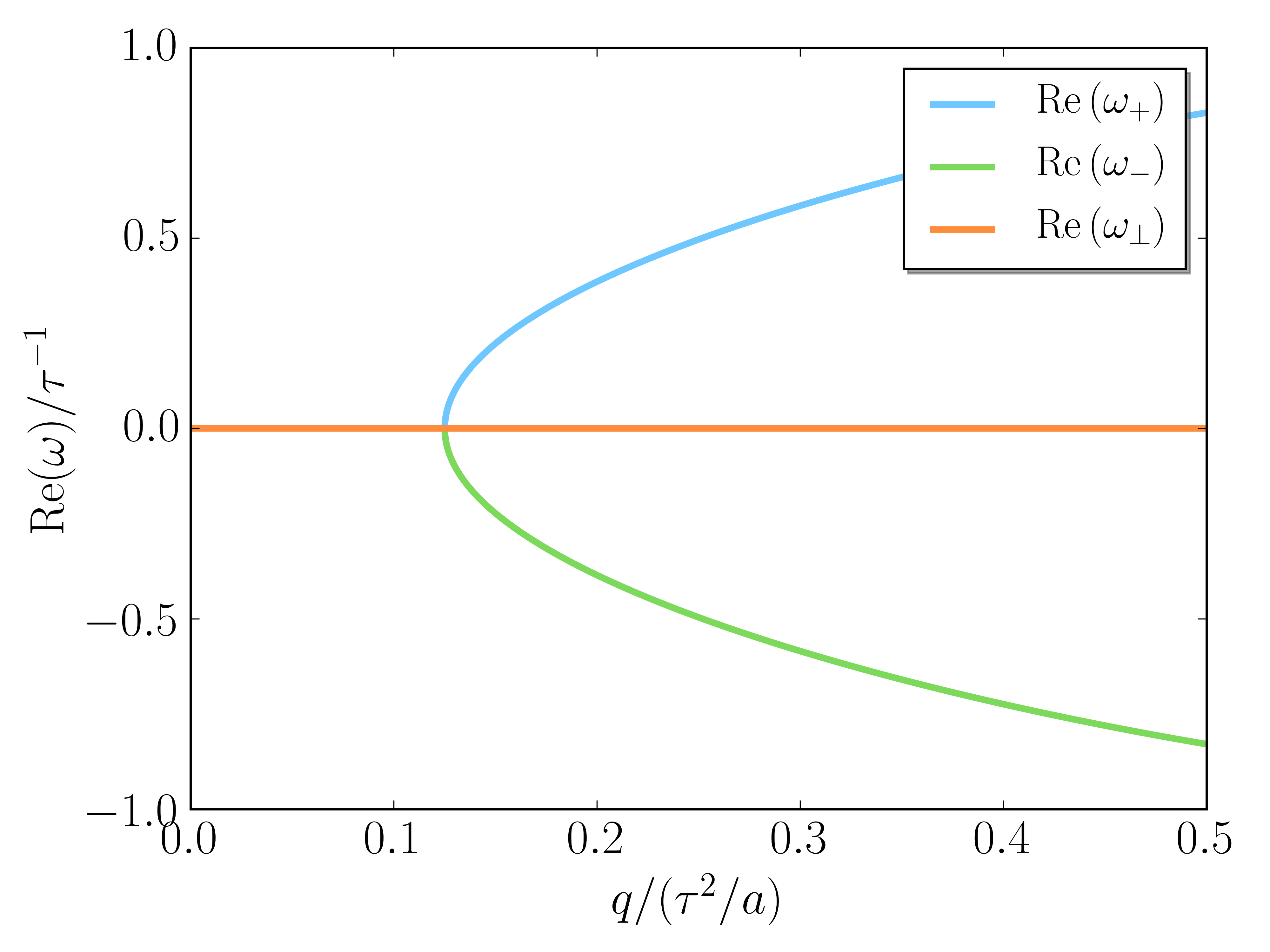}\vfill\includegraphics[scale=0.3]{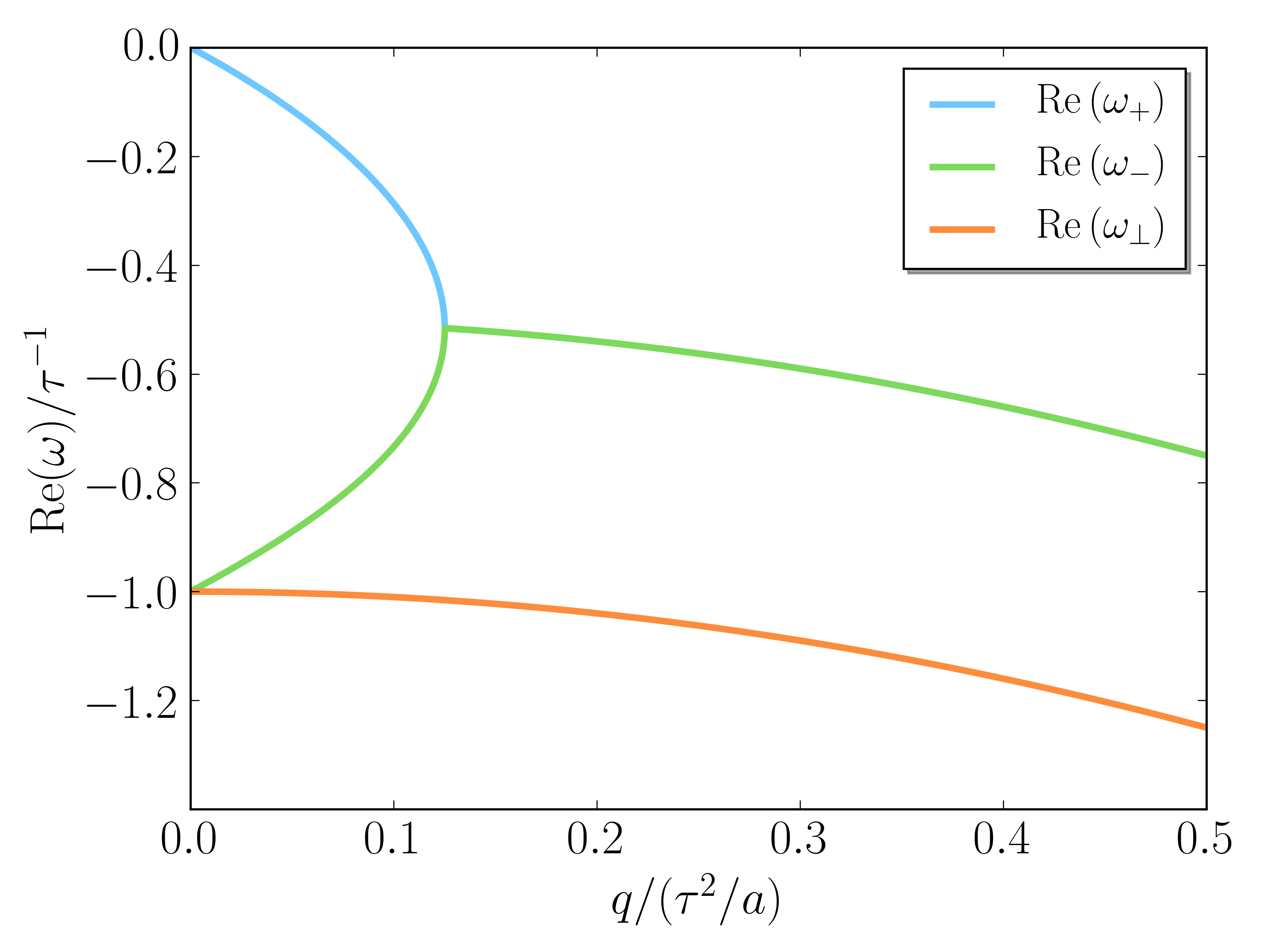}\caption{\label{fig:larger_q_modes}The dispersion relation of the damped plasmon
mode is shown. In the presence of momentum relaxation, the well known
$\sqrt{q}$-plasmon mode is damped out, such that below a certain
threshold wavenumber $\mathrm{Re}\left(\omega_{\pm}\left(q\right)\right)=0$
holds. Interestingly, the plasmon is also damped out at large $q$,
due to both viscosity and momentum relaxation. The blue line corresponds
to the superdiffusive mode $\omega_{+}\left(q\right)$ (see Eq. (\ref{eq:superdiffusive_dispersion})).}
\end{figure}

The full dispersion relations $\omega_{\pm}\left(q\right)$ which
are also valid at larger $q$ can be efficiently obtained from the
well known condition for collective excitations $\varepsilon\left(\omega,\mathbf{q}\right)=0$,
where $\varepsilon\left(\omega,\mathbf{q}\right)=\phi_{\mathrm{ext}}\left(\omega,\mathbf{q}\right)/\phi\left(\omega,\mathbf{q}\right)$
is the dielectric function of the 2D material. Here $\phi=\phi_{\mathrm{ext}}+\phi_{\mathrm{ind}}$
is the total electric potential, where $\phi_{\mathrm{ext}}$ is due
to external sources and $\phi_{\mathrm{ind}}$ is sourced by the inhomogenious
charge carrier density $\rho_{\mathcal{Q}}^{\left(1\right)}$. From
the definition of $\varepsilon$ we find
\begin{equation}
\varepsilon\left(\omega,\mathbf{q}\right)=1-\chi{}_{\rho_{\mathcal{Q}}\rho_{\mathcal{Q}}}\left(\omega,\mathbf{q}\right)V\left(\mathbf{q}\right).\label{eq:dielectric_function}
\end{equation}
Here $\chi{}_{\rho_{\mathcal{Q}}\rho_{\mathcal{Q}}}$ is the charge
susceptibility which is defined via the relation

\begin{equation}
\rho_{\mathcal{Q}}^{\left(1\right)}\left(\omega,\mathbf{q}\right)=\chi{}_{\rho_{\mathcal{Q}}\rho_{\mathcal{Q}}}\left(\omega,\mathbf{q}\right)\phi\left(\omega,\mathbf{q}\right).\label{eq:Charge_suscept_def}
\end{equation}
The condition $\varepsilon\left(\omega,\mathbf{q}\right)=0$ together
with the Eqs. (\ref{eq:longitudinal_conductivity}) and (\ref{eq:Conductivity_suscept_general})
then gives\begin{widetext}
\begin{equation}
\omega_{\pm}\left(q\right)=-i\frac{1}{2\tau}-iq^{2}\frac{\zeta+\eta}{2\rho_{\mathcal{M}}^{\left(0\right)}\tau}\pm\sqrt{2aq\left(1+\frac{Kq}{2a\rho_{\mathcal{M}}^{\left(0\right)}}\right)-\frac{1}{4\tau^{2}}\left(\frac{q^{2}\left(\zeta+\eta\right)}{\rho_{\mathcal{M}}^{\left(0\right)}}+1\right)^{2}}.\label{eq:Larger_q_modes}
\end{equation}
\end{widetext}For small $q$, Eq. (\ref{eq:Larger_q_modes}) reduces
to the expression given in Eq. (\ref{eq:Hydro_dispersion}). A third
mode $\omega_{\perp}$ is easily found by setting $\mathbf{u}\cdot\mathbf{q}=0$
in the Navier-Stokes equation (\ref{eq:Nav_Stokes}). This transverse
mode obeys the dispersion relation
\begin{equation}
\omega_{\perp}=-\frac{i}{\tau}-i\eta q^{2}.\label{eq:Larger_q_transverse}
\end{equation}
The modes of Eqs. (\ref{eq:Larger_q_modes}), (\ref{eq:Larger_q_transverse})
are shown in Fig. \ref{fig:larger_q_modes}.

\subsection{Einstein relation}

Although the charge dynamics in a two dimensional liquid is superdiffusive,
transport coefficients obey the Einstein relation: The longitudinal
conductivity $\sigma_{\parallel}$ is defined via Ohm's law: $j_{\mathcal{Q},i}\left(\omega,\mathbf{q}\right)=\sigma_{\parallel}\left(\omega,\mathbf{q}\right)E_{i}\left(\omega,\mathbf{q}\right)$
with $\mathbf{E}\propto\mathbf{q}$, where the electric field is determined
by the gradient of the total electrostatic potential: $\mathbf{E}=-\nabla\phi$.
In the regime of linear response, we find from Eq. (\ref{eq:Nav_Stokes})
\begin{equation}
\sigma_{\parallel}\left(\omega,\mathbf{q}\right)=\frac{\tau\left(\rho_{\mathcal{Q}}^{\left(0\right)}\right)^{2}/\rho_{\mathcal{M}}^{\left(0\right)}}{1-i\omega\tau+i\tau\frac{q^{2}}{\omega}\frac{K}{\rho_{\mathcal{M}}^{\left(0\right)}}+q^{2}\frac{\eta+\zeta}{\rho_{\mathcal{M}}^{\left(0\right)}}}.\label{eq:longitudinal_conductivity}
\end{equation}
Here we have again used the relation $\nabla p=\left(K/\rho_{\mathcal{Q}}^{\left(0\right)}\right)\nabla\rho_{\mathcal{Q}}^{\left(1\right)}$
where $K=\rho_{\mathcal{Q}}\left(\partial p/\partial\rho_{\mathcal{Q}}\right)$
is the bulk modulus. We also assumed an isotropic system where the
viscous stress tensor reduces to $\tau_{ij}=\eta\partial_{j}\partial_{j}u_{i}+\zeta\partial_{i}\partial_{j}u_{j}$,
with the shear and bulk viscosities $\eta$, $\zeta$. Let us relate
the bulk modulus to the charge susceptibility $\chi{}_{\rho_{\mathcal{Q}}\rho_{\mathcal{Q}}}$.
In the static case $\omega=0$, forces stemming from fluctuations
of $\phi\left(\mathbf{x}\right)$ are balanced by pressure changes:
$-i\mathbf{q}\rho_{\mathcal{Q}}^{\left(0\right)}\phi\left(0,\mathbf{q}\right)=\left(K/\rho_{\mathcal{Q}}^{\left(0\right)}\right)i\mathbf{q}\rho_{\mathcal{Q}}^{\left(1\right)}\left(0,\mathbf{q}\right)$.
It follows
\begin{equation}
\chi{}_{\rho_{\mathcal{Q}}\rho_{\mathcal{Q}}}\left(0,\mathbf{q}\rightarrow0\right)=-\left(\rho_{\mathcal{Q}}^{\left(0\right)}\right)^{2}K^{-1}.\label{eq:Charge_suscept_bulk_modulus}
\end{equation}
This is a special case of the relation
\begin{equation}
\sigma_{\parallel}\left(\omega,\mathbf{q}\right)=\frac{i\omega}{q^{2}}\chi{}_{\rho_{\mathcal{Q}}\rho_{\mathcal{Q}}}\left(\omega,\mathbf{q}\right),\label{eq:Conductivity_suscept_general}
\end{equation}
which can be obtained from the Kubo expression for $\sigma$ and the
continuity equation \cite{PinesNozieres1}. The conductivity (\ref{eq:longitudinal_conductivity})
has a diffusive pole
\begin{equation}
\omega_{D}\left(q\right)=-iDq^{2}\label{eq:Diffusion_pole_Conductivity}
\end{equation}
for small $q$ (see Eq. (\ref{eq:Hydro_dispersion_gated}) for the
full expression). Here, $D$ is the diffusion constant $D=K\tau/\rho_{\mathcal{M}}^{\left(0\right)}$.
However, the pole (\ref{eq:Diffusion_pole_Conductivity}) does not
coincide with the superdiffusive mode $\omega_{+}\left(q\right)$
of Eq. (\ref{eq:superdiffusive_dispersion}), because the electric
conductivity characterizes the system's response to the total electric
field $\mathbf{E}=-\nabla\phi$. For a given electrostatic potential
$\phi$, the charge density is fixed through Eq. (\ref{eq:Charge_suscept_def}).
Indeed, a charge distribution evolving according to Eq. (\ref{eq:Diffusion_pole_Conductivity})
would not solve the Eqs. (\ref{eq:Nav_Stokes}), (\ref{eq:Charge_conti})
(see also Ref. \cite{efros2008_Einstein_Relation} for a similar discussion).
Combining (\ref{eq:longitudinal_conductivity}), (\ref{eq:Charge_suscept_bulk_modulus})
we obtain the Einstein relation
\begin{equation}
D=\frac{\sigma_{\parallel}\left(\omega\rightarrow0,0\right)}{\chi{}_{\rho_{\mathcal{Q}}\rho_{\mathcal{Q}}}\left(0,\mathbf{q}\rightarrow0\right)}.\label{eq:Einstein_diff_constant}
\end{equation}
The order of limits for $\omega$ and $\mathbf{q}$ is essential,
since at finite $\omega$, $\mathbf{q}$, Eq. (\ref{eq:Conductivity_suscept_general})
determines the ratio $\sigma_{\parallel}/\chi{}_{\rho_{\mathcal{Q}}\rho_{\mathcal{Q}}}$.
Finally we note, that even though diffusion is normal for Yukawa interacting
charges (see Eq. (\ref{eq:Yukawa_diffusion_eq})), the diffusion constant
$D_{Y}$ is not equal to the $D$ of Eq. (\ref{eq:Einstein_diff_constant}).

\subsection{Gated systems\label{sub:Gated-systems}}

Two dimensional solid-state systems can be manipulated by gates \cite{Shur2013_GaAs_Devices,Nguyen2019Gating}.
In particular, if the distance between the 2D channel and the gate
is smaller than the length scales of the charge inhomogeneities inside
the 2D layer, the Poisson term (\ref{eq:Poisson_eq}) on the right
of Eq. (\ref{eq:Nav_Stokes}) can be replaced by a capacitive term
$\phi_{C}=\rho_{\mathcal{Q}}^{\left(1\right)}/C$, where $C$ is the
gate capacitance per unit area. This is the local capacitance approximation,
which is appropriate for many gated devices \cite{Dyakonov1993_Dyakonov_Shur_Instability_gated,Dyakonov1996_Nonlinear_Mixing,Torre2019,Zabolotnykh2019_GatedPlasmons}.
The force $-\nabla\phi_{C}$ stemming from the capacitive term can
be absorbed into the pressure term of the Navier-Stokes equation with
the substitution
\[
K\rightarrow\tilde{K}=K+\frac{\rho_{\mathcal{Q}}^{\left(0\right)}}{C}.
\]
The absence of the Poisson term (\ref{eq:Poisson_eq}) changes the
dispersion relations of the hydrodynamic modes. Instead of Eq. (\ref{eq:Omega_characteristic_eq}),
the hydrodynamic modes are now determined by 
\begin{equation}
\omega_{g\pm}=-\frac{i+i\nu q^{2}}{2\tau}\pm\sqrt{\frac{\tilde{K}}{\rho_{\mathcal{M}}^{\left(0\right)}}q^{2}-\frac{\left(1+\nu q^{2}\right)^{2}}{4\tau^{2}}}.\label{eq:Hydro_dispersion_gated}
\end{equation}
The subscript $g$ indicates, that we are considering a gated system
with a uniform electrostatic potential. For small $q$ we have
\begin{eqnarray}
\omega_{g-} & = & -\frac{i}{\tau}-iq^{2}\left(\frac{\nu}{\tau}-\frac{\tilde{K}\tau}{\rho_{\mathcal{M}}^{\left(\text{0}\right)}}\right)\label{eq:gapped_diffusive_mode}\\
\omega_{g+} & = & -\frac{i\tilde{K}q^{2}\tau}{\rho_{\mathcal{M}}^{\left(0\right)}}.\label{eq:diffusive_mode}
\end{eqnarray}
While $\omega_{g-}$ is gapped, $\omega_{g+}$ is an ordinary diffusive
mode with a diffusion constant $\tilde{D}=\tilde{K}\tau/\rho_{\mathcal{M}}^{\left(0\right)}$.
For a gated structure the diffusion constant governing the diffusion
of charges is indeed equal to the one obtained from the Einstein relation
(\ref{eq:Einstein_diff_constant}).

\section{Magnetic fields\label{sec:Magnetic-fields}}

\begin{figure}
\centering{}\includegraphics[scale=0.3]{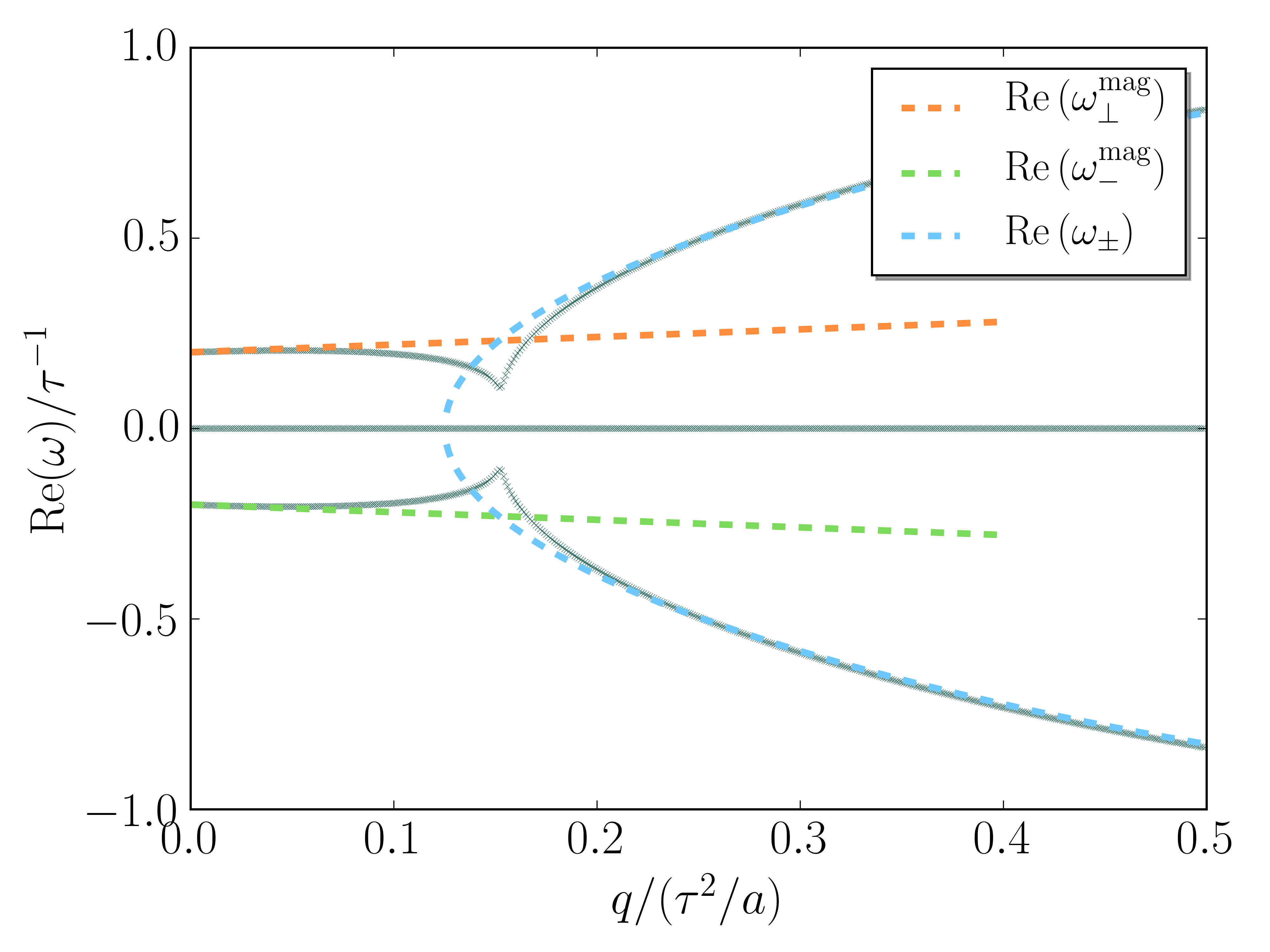}\vfill\includegraphics[scale=0.3]{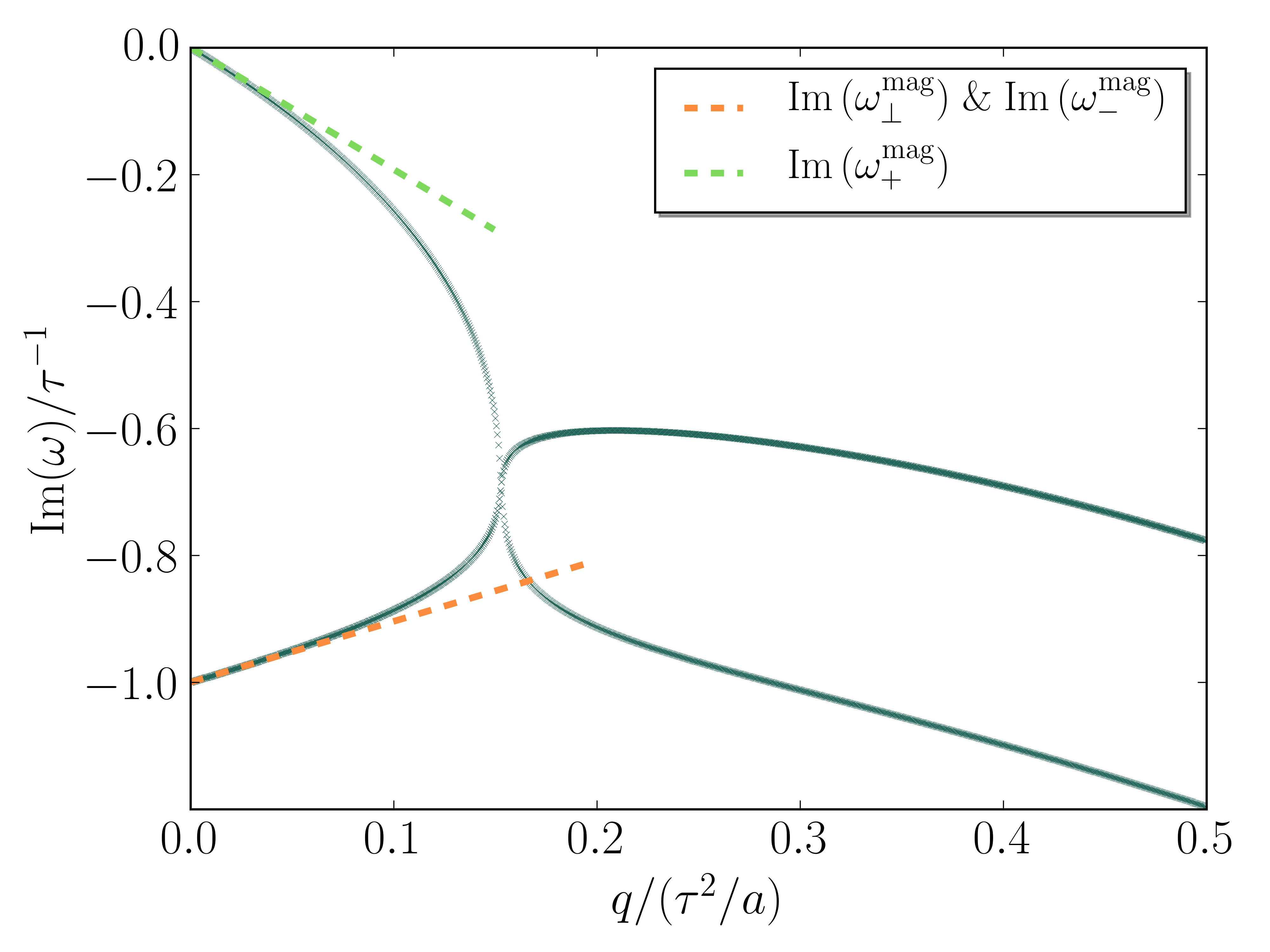}\caption{\label{fig:Mag_0.2_modes}Collective modes of a charged two dimensional
liquid in the presence of momentum relaxation and a perpendicular
magnetic field. The two damped magnetoplasmon modes $\omega_{\perp}^{\mathrm{mag}}$
and $\omega_{-}^{\mathrm{mag}}$ and the superdiffusive mode $\omega_{+}^{\mathrm{mag}}$
are shown. The colored dashed lines correspond to the approximations
of Eqs. (\ref{eq:omega_min_mag}), (\ref{eq:omega_perp_mag}) and
(\ref{eq:omega_pl_mag}). The blue dashed line depicts the damped
plasmon dispersion in the absence of magnetic fields $\omega_{\pm}$
given in Eq. (\ref{eq:Larger_q_modes}), which is a reasonable approximation
to the magnetoplasmon dispersion at larger $q$. At sufficiently large
magnetic fields, the cyclotron resonance and the damped plasmon mode
merge. Here $\omega_{c}=0.2\tau^{-1}$ was chosen, where $\omega_{c}$
is the cyclotron frequency and $\tau^{-1}$ is the rate of momentum
relaxation.}
\end{figure}

\begin{figure}
\centering{}\includegraphics[scale=0.3]{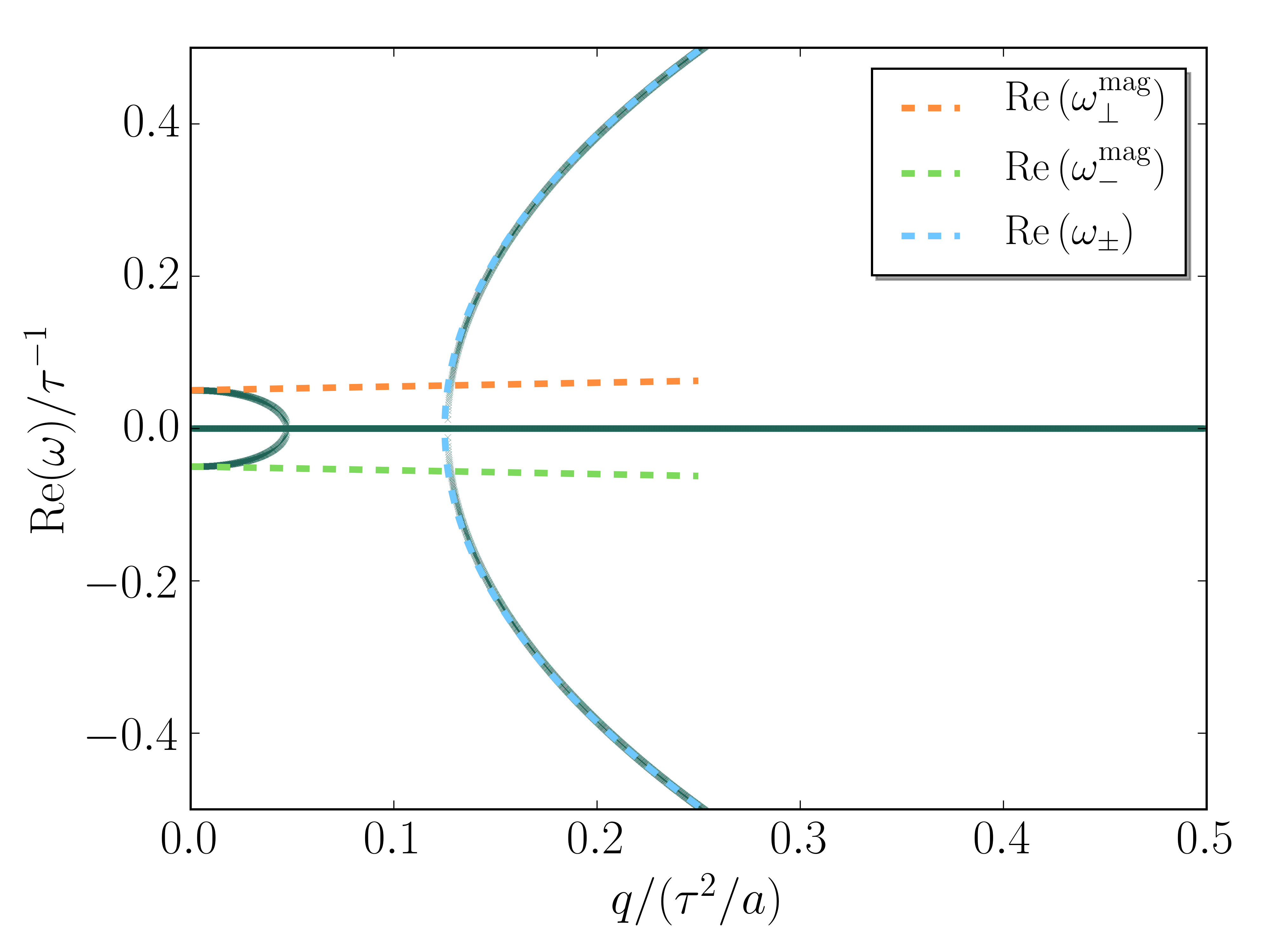}\vfill\includegraphics[scale=0.3]{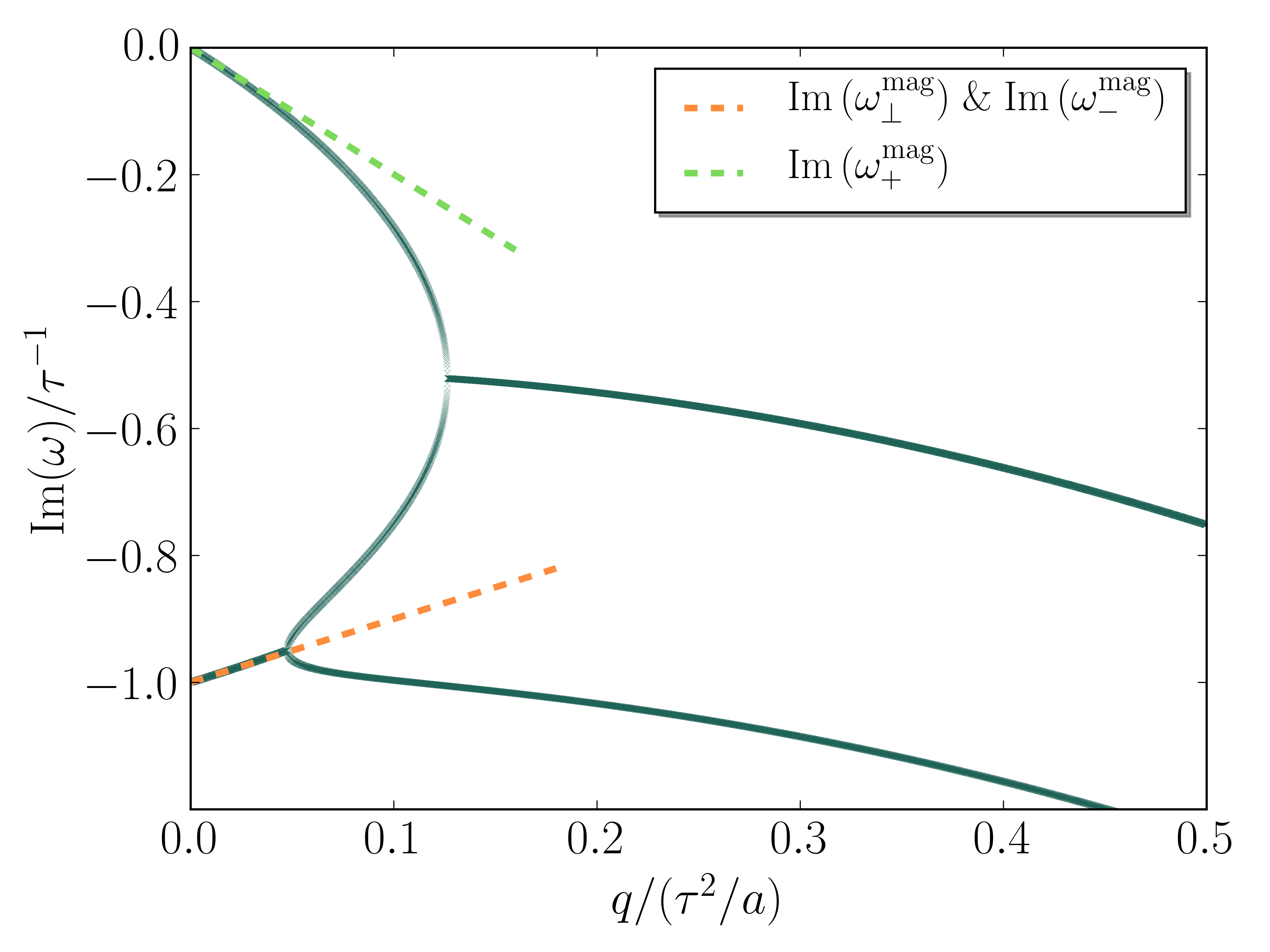}\caption{\label{fig:Mag_0.05_modes}Collective modes of a charged two dimensional
liquid in the presence of momentum relaxation and a perpendicular
magnetic field. In contrast to Fig. \ref{fig:Mag_0.2_modes}, a smaller
magnetic field ($\omega_{c}=0.05\tau^{-1}$) was chosen. The two damped
magnetoplasmon modes $\omega_{\perp}^{\mathrm{mag}}$ and $\omega_{-}^{\mathrm{mag}}$
and the superdiffusive mode $\omega_{+}^{\mathrm{mag}}$ are shown.
The colored dashed lines correspond to the approximations of Eqs.
(\ref{eq:omega_min_mag}), (\ref{eq:omega_perp_mag}) and (\ref{eq:omega_pl_mag}).
The blue dashed line depicts the damped plasmon dispersion in the
absence of magnetic fields $\omega_{\pm}$ given in Eq. (\ref{eq:Larger_q_modes}),
which is a reasonable approximation to the magnetoplasmon dispersion
at larger $q$. The cyclotron resonance and the damped plasmon mode
are well separated, whereas at larger fields strengths, these modes
merge (see Fig. \ref{fig:Mag_0.2_modes}).}
\end{figure}

Magnetic fields qualitatively change the spectrum of collective excitations
of a charged liquid. Under the influence of a magnetic field $B$
charges oscillate at the cyclotron frequency 
\begin{equation}
\omega_{c}=B\rho_{\mathcal{Q}}^{\left(0\right)}/\rho_{\mathcal{M}}^{\left(0\right)}c.
\end{equation}
At finite wavevectors, the cyclotron resonance merges with the plasmon
and gives rise to the magnetoplasmon mode. We are interested in how
the dispersion relations of collectives modes change when both a momentum
relaxation time $\tau$ and a uniform magnetic field are added. This
can be studied by adding a uniform magnetic field oriented perpendicular
to the fluid plane to the Navier-Stokes equation:
\begin{equation}
\partial_{t}\left(\rho_{\mathcal{M}}u_{i}\right)+\partial_{j}\Pi_{ij}=-\frac{1}{\tau}\rho_{\mathcal{M}}u_{i}-\rho_{\mathcal{Q}}\nabla\phi+\frac{B}{c}\varepsilon_{ij}j_{\mathcal{Q},j}.\label{eq:Magnetic_Nav_Stokes}
\end{equation}
The last term in Eq. (\ref{eq:Magnetic_Nav_Stokes}) describes the
Lorentz force exerted on the fluid by the magnetic field. We assume
that the magnetic field is weak, such that Landau quantizuation effects,
as well as the localization of electrons on cyclotron orbits can be
neglected. Linearizing Eq. (\ref{eq:Magnetic_Nav_Stokes}) in $u_{i}$,
$\rho_{\mathcal{M}}^{\left(1\right)}$ and $\rho_{\mathcal{Q}}^{\left(1\right)}$
, performing a Fourier transform and writing the equation in terms
of matrices, we find
\begin{equation}
\mathcal{D}\left(\omega,\mathbf{q}\right)\mathbf{u}\left(\omega,\mathbf{q}\right)=0,
\end{equation}
with\begin{widetext}
\begin{equation}
\mathcal{D}=\left[\begin{array}{cc}
\left(-i\omega+\tau^{-1}\right)+\left(\frac{2ai}{\omega q}+\frac{iK}{\rho_{\mathcal{M}}^{\left(0\right)}\omega}+\frac{\zeta}{\rho_{\mathcal{M}}^{\left(0\right)}}\right)q_{1}^{2}+\nu q^{2} & \left(\frac{2ai}{\omega q}+\frac{iK}{\rho_{\mathcal{M}}^{\left(0\right)}\omega}+\frac{\zeta}{\rho_{\mathcal{M}}^{\left(0\right)}}\right)q_{1}q_{2}-\omega_{c}\\
\left(\frac{2ai}{\omega q}+\frac{iK}{\rho_{\mathcal{M}}^{\left(0\right)}\omega}+\frac{\zeta}{\rho_{\mathcal{M}}^{\left(0\right)}}\right)q_{1}q_{2}+\omega_{c} & \left(-i\omega+\tau^{-1}\right)+\left(\frac{2ai}{\omega q}+\frac{iK}{\rho_{\mathcal{M}}^{\left(0\right)}\omega}+\frac{\zeta}{\rho_{\mathcal{M}}^{\left(0\right)}}\right)q_{2}^{2}+\frac{\eta}{\rho_{\mathcal{M}}^{\left(0\right)}}q^{2}
\end{array}\right].
\end{equation}
Here $\nu=\eta/\rho_{\mathcal{M}}^{\left(0\right)}$ is the kinematic
viscosity. The dispersion relations of collective modes can be found
by setting
\begin{equation}
\det\left(\mathcal{A}\right)=0.\label{eq:det_cond}
\end{equation}
In the limit $\tau\rightarrow\infty$, Eq. (\ref{eq:det_cond}) gives
the magnetoplasmon dispersion \cite{Horing1976_Magnetoplasmon,Mast1985_Magnetoplasmon}
\begin{equation}
\omega_{mp,\pm}=\pm\sqrt{2aq+\omega_{c}^{2}},\label{eq:magnetoplasmon}
\end{equation}
where the conventional square-root plasmon spectrum is gapped out
by the magnetic field. It is interesting to note that in Eq. (\ref{eq:magnetoplasmon})
the two limits $q\rightarrow0$ and $\omega_{c}^{2}\rightarrow0$
are not interchangeable, yielding 
\begin{equation}
\omega_{mp,+}\approx\sqrt{2aq}+\frac{\omega_{c}^{2}}{2\sqrt{2aq}}\,,\quad\omega_{c}^{2}\ll aq
\end{equation}
and 
\begin{equation}
\omega_{mp,+}\approx\omega_{c}+aq/\omega_{c}\,,\quad aq\ll\omega_{c}^{2}.
\end{equation}
Either the cyclotron motion or the plasmon waves dominate the collective
behavior. This behavior is even more striking at finite $\tau$. The
modes $\omega_{-}$, $\omega_{\perp}$ of the Eqs. (\ref{eq:Hydro_dispersion})
and (\ref{eq:Transverse_mode}) then become
\begin{eqnarray}
\omega_{-}^{\mathrm{mag}} & \approx & -\frac{i}{\tau}-ic_{1}+2ia\tau\left(1-\omega_{c}^{2}\tau^{2}\right)q-ic_{2}q-i\frac{c_{3}}{q}\,,\quad\omega_{c}^{2}\ll aq\label{eq:omega_min_1stLim}\\
\omega_{-}^{\mathrm{mag}} & \approx & -\omega_{c}\left(1+a\tau^{2}q\right)-\frac{i}{\tau}+ia\tau\left(1-\omega_{c}^{2}\tau^{2}\right)q\,,\quad aq\ll\omega_{c}^{2}\label{eq:omega_min_mag}\\
\omega_{\perp}^{\mathrm{mag}} & \approx & -\frac{i}{\tau}+ic_{1}+ic_{2}q+i\frac{c_{3}}{q}\,,\quad\omega_{c}^{2}\ll aq\label{eq:omega_perp_1stLim}\\
\omega_{\perp}^{\mathrm{mag}} & \approx & \omega_{c}\left(1+a\tau^{2}q\right)-\frac{i}{\tau}+ia\tau\left(1-\omega_{c}^{2}\tau^{2}\right)q\,,\quad aq\ll\omega_{c}^{2},\label{eq:omega_perp_mag}
\end{eqnarray}
where we have used the abbreviations $c_{1}=\omega_{c}^{2}\frac{\zeta-K\tau}{4a^{2}\tau^{2}\rho_{\mathcal{M}}^{\left(0\right)}}$,
$c_{2}=\omega_{c}^{2}\frac{4a^{2}\tau^{3}\nu\rho_{\mathcal{M}}^{\left(0\right)}+\tau^{2}K^{2}+\zeta^{2}-2\zeta K\tau}{8a^{3}\tau^{3}\left(\rho_{\mathcal{M}}^{\left(0\right)}\right)^{2}}$
and $c_{3}=\frac{\omega_{c}^{2}}{2a\tau}$. In the limit of small
but finite magnetic fields, the modes $\omega_{-}^{\mathrm{mag}}$,
$\omega_{\perp}^{\mathrm{mag}}$ acquire a dispersive real part of
$\pm\omega_{c}\left(1+a\tau^{2}q\right)$ and become wavelike, albeit
heavily damped. The mode spectrum for larger $q$ is quite complicated
and is depicted in Figs. \ref{fig:Mag_0.2_modes} and \ref{fig:Mag_0.05_modes}.
The noninterchangeability of the $q\rightarrow0$ and $\omega_{c}^{2}\rightarrow0$
limits is a subtle issue since both variables set length scales that
influence the transport behavior: either magnetotransport or nonlocal
effects dominate. Plasmon spectra for different values of $\omega_{c}$
and $\tau$ are the subject of Ref. \cite{Volkov2016magnetoplasmon_retardation_effects}.
The authors point out, that electrodynamic retardation effects may
play an important role. For 2D undamped plasmons in the absence of
magnetic fields retardation becomes important for $q\rightarrow0$,
when the phase velocity approaches the the speed of light $c$. In
the case of damped magnetoplasmons, the interplay between $\omega_{c}\tau$,
$q$ and the dc conductivity $\sigma$ which can approach $c$ becomes
important. However, the superdiffusive mode, which is under consideration
here, is save from these effects. For the superdiffusive mode (\ref{eq:superdiffusive_dispersion})
the two limits $q\rightarrow0$ and $\omega_{c}^{2}\rightarrow0$
are interchageable. In both cases the superdiffusive dispersion reads
\begin{equation}
\omega_{+}^{\mathrm{mag}}=-2ia\tau\left(1-\omega_{c}^{2}\tau^{2}\right)q.\label{eq:omega_pl_mag}
\end{equation}
The superdiffusion is thus slower by a factor of $1-\omega_{c}^{2}\tau^{2}$
for $\omega_{c}\tau\ll1$.\end{widetext}


\section{Langevin equations and Lévy flights\label{sec:Langevin-equations} }

Diffusion processes can be modeled with Langevin-type stochastic equations.
Here, we demonstrate that charged particles interacting via the Coulomb
potential, while undergoing diffusion, indeed follow Lévy flight trajectories
that produce the superdiffusive dynamics of Eq. (\ref{eq:superdiffusive_dispersion}).
Our starting point are coupled Langevin equations for the particle
coordinates $\mathbf{r}_{i}\left(t\right)$ \cite{Chandrasekhar1960,Ott2014_One_component_Coulomb,Ott2009_Diffusion}:
\begin{equation}
\ddot{\mathbf{r}}^{\left(i\right)}=-\frac{\mathcal{Q}^{2}}{m}\sum_{j\neq i}^{N}\frac{1}{\left|\mathbf{r}^{\left(j\right)}-\mathbf{r}^{\left(i\right)}\right|}-\frac{1}{\tau}\dot{\mathbf{r}}+\frac{1}{m}\boldsymbol{\eta}^{\left(i\right)}.\label{eq:Langevin_Eq}
\end{equation}
$\mathcal{Q}$ is the particle charge and $\boldsymbol{\eta}_{i}$
is an uncorrelated stochastic force for which holds $\left\langle \eta_{k}^{\left(i\right)}\left(t\right)\eta_{l}^{\left(j\right)}\left(t'\right)\right\rangle =\kappa\delta_{ij}\delta_{kl}\delta\left(t-t'\right).$
The Eq. (\ref{eq:Langevin_Eq}) could describe a one-component Coulomb
plasma \cite{Ott2014_One_component_Coulomb} (as e.g. realized by
macroions in colloidal suspensions \cite{Uuguz2009_Colloidal_Suspensions,Warren2000_Colloidal_Suspensions}).
Here $\kappa=2mk_{B}T\tau^{-1}$ holds due to the Einstein relation. 

It is a textbook result that without the Coulomb term in Eq. (\ref{eq:Langevin_Eq}),
the particles will undergo Brownian motion. Indeed, for $\mathcal{Q}=0$
the response to the stochastic force is given by 
\[
\mathbf{r}^{\left(i\right)}\left(t\right)=\frac{\tau}{m}\int_{0}^{t}dt'\,\left(1-e^{-\frac{\left(t-t'\right)}{\tau}}\right)\boldsymbol{\eta}^{\left(i\right)}\left(t'\right)
\]
where we assumed that $\boldsymbol{\eta}$ was switched on at $t=0$.
For the variance $\mathbf{r}^{\left(i\right)}\left(t\right)$ follows
\[
\left\langle \mathbf{r}^{\left(i\right)}\left(t\right)\cdot\mathbf{r}^{\left(j\right)}\left(t\right)\right\rangle \approx\delta_{ij}\frac{\tau^{2}\kappa}{m^{2}}t,
\]
which corresponds to an ordinary Gaussian diffusion process. To demonstrate,
how the non-Gaussian superdiffusive dynamics of Eq. (\ref{eq:superdiffusive_dispersion})
emerges once the Coulomb interactions are turned on, we integrate
Eq. (\ref{eq:Langevin_Eq}) numerically. In the simulations, we used
periodic boundary conditions and have chosen $t_{c}=\sqrt{mr_{c}^{3}/\mathcal{Q}^{2}}$
as our unit of time. The length $r_{c}$ is chosen arbitrarily (box
size $L=25r_{c}$) but can be related to the Wigner-Seitz radius $a$:
$a=\sqrt{\pi n}\approx1.54r_{c}$. At $t=0$, the system consists
of a uniform background distribution of particles $\rho_{\mathcal{Q}}^{\left(0\right)}=0.72/r_{c}^{2}$
and a small number of non-equlibrium particles ($n=20$) localized
completely within the unit square $\Theta\left(1-2\left|x\right|\right)\Theta\left(1-2\left|y\right|\right)$.
The coupling parameter $\Gamma$ is small $\Gamma=\mathcal{Q}^{2}/\left(ak_{B}T\right)\approx0.05$
and the damping is substantial: $\tau=0.1t_{c}$.

The movements of the particles at $t>0$ can be interpreted as random
walks. From the discussion of Sec. \ref{sec:Superdiffusion}, one
expects that the distances travelled by the particles during an interval
$\Delta t$ are distributed according to the Cauchy distribution Eq.
(\ref{eq:cauchy_solution}). I.e., if the diffusion is indeed anomalous
with a coefficient $\alpha=1$, the step sizes $\Delta r$ of the
random walk will follow a fat-tail powerlaw distribution decaying
as 
\begin{equation}
p\left(\Delta r\right)\sim\frac{1}{\Delta r^{3}}.\label{eq:step_sizes_tail}
\end{equation}
The superdiffusive behavior manifests itself at small wavevectors,
i.e. large distances, therefore the behavior for small step sizes
will deviate from the $\Delta r^{-3}$ law. Since the variance of
the Cauchy distribution is not defined, we cannot identify the superdiffusive
dynamics by measuring the correlation function $\left\langle \mathbf{r}^{\left(i\right)}\left(t\right)\cdot\mathbf{r}^{\left(j\right)}\left(t\right)\right\rangle $.
Instead, the step size distribution of Eq. (\ref{eq:step_sizes_tail})
can be used to study the Lévy flight nature of the diffusion process.

Fig. \ref{fig:Langevin_x^=00007B-3=00007D_tail} shows the step size
distribution obtained in our computational experiment. The power-law
decay of Eqs. (\ref{eq:cauchy_solution}), (\ref{eq:step_sizes_tail})
for large $\Delta r$ is clearly observed. Thus, the diffusive random
motion of Coulomb interacting two dimensional particles is a non-Gaussian,
Lévy stable random walk. From the general properties of such random
walks, we know that the mean travelled distance of a particles grows
as $t$, in contrast to the $\sqrt{t}$ scaling of normal diffusion
\cite{Bouchaud1990,Gnedenko1954}.

\section{Specific heat\label{sec:Specific-heat}}

Finally, we want to discuss the contributions of the (super)diffusive
modes to the specific heat $C$ of a charged two dimensional liquid.
The thermodynamics of collective excitation has been the subject of
many studies (see e.g. \cite{Hoepfel1982_Thermal_plasmons,Trachenko2015collective_thermodynamics,Baggioli2019BosonPeak_diffusive,Baggioli2019LinearTCv_diffusion}).
In Appendix \ref{app:Specific-Heat-Contribution}, we show that the
modes' contribution to internal energy density $E$ is given by 
\[
E=\int_{0}^{\infty}d\varepsilon\frac{\varepsilon\nu\left(\varepsilon\right)}{e^{\beta\varepsilon}-1},
\]
where $\beta=1/k_{B}T$ and $\nu\left(\varepsilon\right)$ is the
density of states of the modes:
\begin{equation}
\nu\left(\varepsilon\right)=-\frac{1}{\pi}\int_{0}^{q^{*}}\frac{qdq}{2\pi}\mathrm{Im}\left[G\left(\varepsilon,\mathbf{q}\right)\right].\label{eq:Superdiff_DOS}
\end{equation}
$G\left(\varepsilon,\mathbf{q}\right)$ is the Green's function of
the diffusion equation. $q^{*}$ serves as a momentum cut-off. The
specific heat is defined as
\[
c_{V}=\frac{\partial E}{\partial T}.
\]
To simplify the analysis, we will focus on low temperatures, where
the relevant modes will be the superdiffusive mode of Eq. (\ref{eq:superdiffusive_dispersion})
while the gapped mode (\ref{eq:gapped_decaying_dispersion}) will
only gain importance at higher temperatures. We begin with the superdiffusive
mode $\omega_{+}=-2ia\left|q\right|$ and obtain
\begin{equation}
\nu\left(\varepsilon\right)=\frac{2aq^{*}\tau-\varepsilon\tan^{-1}\left(2aq^{*}\tau/\varepsilon\right)}{8a^{2}\tau^{2}\pi^{2}}.\label{eq:Superdiff_DOS_result}
\end{equation}
The superdiffusive mode contribution to the heat capacity for small
temperatures is then given by 
\[
c_{V}=c_{1}T-c_{2}T^{2}+c_{3}T^{3}-\mathcal{O}\left(T^{5}\right),
\]
where all higher orders are of odd powers in $T$. For the coefficient
of the linear term we obtain $c_{1}=\frac{q^{*}}{24a\tau}$. Surprisingly,
$c_{2}=\frac{3\zeta(3)}{8\pi a^{2}\tau^{2}}$ does not depend on the
momentum cut-off $q^{*}$ and has a negative sign (although $C$ is
always positive). This is due to the fact that the $T^{2}$-dependence
can be traced back to the nonanalyticity of the superdiffusive mode
at small $q$. We observe that the integrand of Eq. (\ref{eq:Superdiff_DOS})
depends on $\varepsilon^{2}$, yet the density of states $\nu\left(\varepsilon\right)$
has a linear term in $\varepsilon$. We can extract this term from
the integral of Eq. (\ref{eq:Superdiff_DOS}): 
\begin{eqnarray}
\lim_{\varepsilon\rightarrow0}\frac{\nu\left(\varepsilon\right)-\nu\left(0\right)}{\varepsilon} & = & \lim_{\varepsilon\rightarrow0}\frac{1}{2\pi a\tau}\int_{0}^{q^{*}}\frac{dq}{2\pi}\frac{1}{\varepsilon}\left(\frac{4a^{2}\tau^{2}q^{2}}{\varepsilon^{2}+4a^{2}\tau^{2}q^{2}}-q\right)\nonumber \\
 & = & -\frac{1}{16\pi a^{2}\tau^{2}}\int_{-\infty}^{\infty}dq\,\delta\left(q\right).\label{eq:non-analyt_t^2}
\end{eqnarray}
For small energies, the DOS is 
\begin{equation}
\nu\left(\varepsilon\right)\approx\frac{q^{*}}{4\pi^{2}a\tau}-\frac{\varepsilon}{16\pi a^{2}\tau^{2}},
\end{equation}
giving the above values of $c_{1}$, $c_{2}$. It is well known that
to leading order at low temperatures, the specific heat $C$ of Galilei
invariant Fermi liquids is linear in the temperature, just as for
free Fermions. However the analogy does not hold beyond the leading
order term. Nonanalytic terms in the fermion self energy of two dimensional
Fermi liquids result in corrections to the specific heat $\delta C$
which behave as $\delta C\sim T^{2}$ \cite{Belitz1997Nonanalytic_FL,Chubukov2003Nonanalytic_FL,Chubukov2004Singular_FL}.
This result is true for both Coulomb and short range interactions
\cite{Chubukov2005_Singular_2D}. Here we show that in the presence
of momentum relaxation, the nonanalytic superdiffusive mode of Eq.
(\ref{eq:superdiffusive_dispersion}) as well contributes a $\sim T^{2}$
correction to the specific heat, however with an opposite sign. This
is in contrast to the plasmon resonance of a two dimensional charged
system where the plasmon dispersion is given by $\omega=\sqrt{2aq}$
and only contributes a sub-subleading $\sim T^{4}$ term \cite{Hoepfel1982_Thermal_plasmons}.

For the specific heat contribution of the diffusive mode $\omega_{g+}$
of gated 2D systems we obtain
\begin{equation}
c_{V,g}=d_{1}T+d_{1}'T\log\left(\frac{1}{T}\right)+d_{3}T^{3}+\mathcal{O}\left(T^{5}\right).\label{eq:Diffusive_SH_contr}
\end{equation}
Gating thus qualitatively changes the specific heat of a charged two
dimensional system. The low temperature behavior will be dominated
by the $T\log\left(1/T\right)$ term. Interestingly, several mechanisms
have been discussed that lead to a $T\log\left(1/T\right)$ temperature
dependence of the specific heat in two dimensional systems, such as
quantum critical fluctuations of overdamped bosonic modes with a dynamical
exponent $z=2$ \cite{Millis1993_Temperature_quantum_critical} and
scattering between hot Fermi pocket and cold Fermi surface electrons
in Sr$_{3}$Ru$_{2}$O$_{7}$ \cite{Mousatov2020_strange_metal_Sr3Ru2O7}.
For Sr$_{3}$Ru$_{2}$O$_{7}$, the $T\log\left(1/T\right)$ contribution
has been observed experimentally \cite{Sun2018_SpecificHeat_Sr3Ru2O7}.

\section{Acknowedgements}

I am greatful to Michael Bonitz, Igor Gornyi and Jörg Schmalian for
helpful comments and discussions. I also want thank Bhilahari Jeevanesan
and Jonas Karcher who helped to improve this manuscript.

\begin{appendix}

\section{Charge and mass densities in systems without Galilean invariance\label{app:Charge-and-Mass}}

In Galilean invariant systems the notions of mass and charge densities
are straightforward. If $\rho$ is the particle number density, the
mass density is given by $\rho_{\mathcal{M}}=m\rho$ and the charge
density by $\rho_{\mathcal{Q}}=e\rho$, where $m$ and $e$ are the
mass and charge of a particle. However many solid state systems do
not exhibit Galilean invariance, and it is usefull to extend the definitions
of mass and charge densities to non-Galilean invariant, yet translation
invariant systems, where momentum conservation ensures the validity
of hydrodynamics. Here the velocity $u_{i}\left(\mathbf{x}\right)$
is defined as a source of the conserved crystal momentum \cite{lucas2015memory}.
With the shift 
\begin{equation}
H\rightarrow H-\int d^{2}x\,u_{i}\left(t,\mathbf{x}\right)P_{i}\left(\mathbf{x}\right),\label{eq:Velocity_source_momentum}
\end{equation}
where $H$ is the full Hamiltonian of the system and $P_{i}\left(\mathbf{x}\right)$
is the momentum operator, the densities $\rho_{\mathcal{M}}$$\left(t,\mathbf{x}\right)$,
$\rho_{\mathcal{Q}}\left(t,\mathbf{x}\right)$ can be defined as response
functions and calculated using the memory matrix formalism \cite{Forster,lucas2015memory}.
Memory matrices allow to construct a hydrodynamic approximation to
a quantum system by restricting the infinite-dimensional space of
possible observables to a few conserved quantities and quantities
which decay at very long time scales. Sticking to the notation of
Ref. \cite{lucas2015memory}, we will call these quantities $X_{A}$.
Their thermodynamic conjugate shall be called $U_{B}$. An important
object is the generalized conductivity $\sigma_{AB}$. The memory
matrix formalism provides efficient means for its calculation. The
generalized conductivity relates the quantities $X_{A}$ to the fields
$\dot{U}_{B}$: 
\begin{equation}
\left\langle X_{A}\right\rangle =-\sigma_{AB}\dot{U}_{B},\label{eq:Generalized_Conductivity_Ohm}
\end{equation}
where a summation over the index $B$ labelling the (quasi)conserved
quantities is implied. $\sigma_{AB}$ can be expressed in terms of
retarded Green's functions 
\begin{equation}
\sigma_{AB}\left(z,\mathbf{q}\right)=\frac{1}{iz}\left(G_{AB}^{R}\left(z,\mathbf{q}\right)-G_{AB}^{R}\left(i0,\mathbf{q}\right)\right)\label{eq:Gen_cond_Greens}
\end{equation}
 with 
\[
G_{AB}^{R}\left(t,\mathbf{x}\right)=-i\Theta\left(t\right)\left\langle \left[X_{A}\left(t,\mathbf{x}\right),X_{B}\left(0,\mathbf{0}\right)\right]\right\rangle .
\]
As is customary in memory matrix literature, we used the Laplace transform
\[
G_{AB}^{R}\left(z,\mathbf{q}\right)=\int_{0}^{\infty}dt\,e^{izt}G_{AB}^{R}\left(t,\mathbf{q}\right).
\]

In our case the quantities $X_{A}$ include the momentum $P_{i}$,
which, following Eq. (\ref{eq:Velocity_source_momentum}), is sourced
by the velocity $u_{i}$. Using Eq. (\ref{eq:Generalized_Conductivity_Ohm}),
we write
\begin{equation}
\left\langle P_{i}\right\rangle =-iz\sigma_{P_{i},u_{i}}u_{i}.
\end{equation}
The above equation suggests that the mass density should be defined
as $\rho_{\mathcal{M}}=-iz\sigma_{P_{i},u_{i}}$. It follows
\begin{equation}
\left\langle P_{i}\right\rangle \left(t,\mathbf{x}\right)=\int dt'd^{2}x'\,\rho_{\mathcal{M}}\left(t-t',\mathbf{x}-\mathbf{x}'\right)u_{i}\left(t',\mathbf{x}'\right).\label{eq:nonlocal_mass_density}
\end{equation}
Finally, we should keep in mind that the scales of hydrodynamic temporal
and spatial inhomogeneities $t_{\mathrm{h}}$ and $l_{\mathrm{h}}$
are much smaller than any time or length scale $\tilde{t}$, $\tilde{l}$
characterizing the Hamiltonian $H$. Since the Green's functions in
the Eq. (\ref{eq:Gen_cond_Greens}) are calculated at vanishing flow
velocities, they will decay on scales given by $\left|t-t'\right|\approx\tilde{t}$,
$\left|\mathbf{x}-\mathbf{x}'\right|\approx\tilde{l}$. On the other
hand, the flow velocity $u_{i}$ varies on scales $t_{\mathrm{h}}$,
$l_{\mathrm{h}}$. Thus, in the hydrodynamic limit $\tilde{t}\ll t_{h}$,
$\tilde{l}\ll l_{h}$, Eq. (\ref{eq:nonlocal_mass_density}) can be
approximated by the local relation
\begin{equation}
g_{i}\left(t,\mathbf{x}\right)\approx\rho_{\mathcal{M}}\left(t,\mathbf{x}\right)u_{i}\left(t,\mathbf{x}\right).\label{eq:Mass_density_def}
\end{equation}
$\rho_{\mathcal{M}}\left(t,\mathbf{x}\right)$ is the mass density
used throughout the text. Similarely, we arrive at $\rho_{\mathcal{Q}}=-iz\sigma_{J_{\mathcal{Q},i},u_{i}}$,
where $J_{\mathcal{Q},i}$ is the electric current operator, and finally
\begin{equation}
j_{\mathcal{Q},i}\approx\rho_{\mathcal{Q}}\left(t,\mathbf{x}\right)u_{i}\left(t,\mathbf{x}\right).\label{eq:Charge_density_def}
\end{equation}

\section{Specific heat contribution of collective modes\label{app:Specific-Heat-Contribution}}

We begin with the partition function
\begin{equation}
Z=\int\mathcal{D}\left[\phi_{\mathbf{q}}\right]e^{-\sum_{\mathbf{q},n}\phi_{\mathbf{q},i\omega_{n}}G^{-1}\left(i\omega_{n},\mathbf{q}\right)\phi_{\mathbf{q},i\omega_{n}}}.
\end{equation}
Here, $G\left(i\omega_{n},\mathbf{q}\right)$ is the Green's function
of the damped bosonic plasmon mode
\begin{equation}
G\left(i\omega_{n},\mathbf{q}\right)=\frac{1}{-i\omega_{n}+\omega_{+}\left(\mathbf{q}\right)},
\end{equation}
where $\omega_{+}\left(\mathbf{q}\right)$ was introduced in Eq. (\ref{eq:superdiffusive_dispersion}).
$\phi_{\mathbf{q}}$ is representing the bosonic plasmon fields. The
heat capacity can be calculated from the internal energy $E$, which
is given by \cite{AltlandSimons,negele2018quantum}
\begin{equation}
E=-\frac{\partial}{\partial\beta}\int\frac{d^{2}q}{\left(2\pi\right)^{2}}\sum_{n}\ln\left[\beta G^{-1}\left(i\omega_{n},\mathbf{q}\right)\right].
\end{equation}
First, we evaluate the Matsubara sum over bosonic frequencies $\omega_{n}=2\pi n/\beta$
by rewriting it as a contour integral around the imaginary axis of
a variable $\varepsilon$:
\begin{eqnarray*}
E & = & -\frac{\partial}{\partial\beta}\int\frac{d^{2}q}{\left(2\pi\right)^{2}}\int_{\mathcal{C}}\frac{d\varepsilon}{\left(2\pi i\right)}\frac{\beta}{e^{\beta\varepsilon}-1}\ln\left[\beta\left(-\varepsilon+\omega_{+}\left(\mathbf{q}\right)\right)\right].
\end{eqnarray*}
In the following, it will be convenient to use the abbreviation 
\begin{equation}
\omega_{+}\left(\mathbf{q}\right)\approx-2ia\tau_{1}q\equiv\xi.
\end{equation}
The integrand has a branch cut at $\mathrm{Re}\varepsilon>0$ and
$\mathrm{Im}\varepsilon=\xi$. Correspondingly, the contour can be
deformed such that it encircles the line $\mathrm{Im}\varepsilon=-2ia\tau_{1}q$
running from $0$ to infinity and back. The internal energy is then
given by\begin{widetext}
\begin{eqnarray}
E & = & -\frac{\partial}{\partial\beta}\int\frac{d^{2}q}{\left(2\pi\right)^{2}}\int_{0}^{\infty}\frac{d\varepsilon}{\left(2\pi i\right)}\left\{ \frac{\beta}{e^{\beta\left(\varepsilon+\xi\right)}-1}\ln\left[\beta\left(-\left(\varepsilon+\xi+i0^{+}\right)+\xi\right)\right]-\frac{\beta}{e^{\beta\left(\varepsilon+\xi\right)}-1}\ln\left[\beta\left(-\left(\varepsilon+\xi-i0^{+}\right)+\xi\right)\right]\right\} \nonumber \\
\end{eqnarray}
\end{widetext}or
\begin{equation}
E=-\frac{\partial}{\partial\beta}\int\frac{d^{2}q}{\left(2\pi\right)^{2}}\int_{0}^{\infty}\frac{d\varepsilon}{\pi}\frac{\beta}{e^{\beta\varepsilon}-1}\mathrm{Im}\left(\ln\left[\beta\left(-\varepsilon+\xi\right)\right]\right).
\end{equation}
Simplifying the expression we obtain
\begin{eqnarray}
E & = & -\frac{\partial}{\partial\beta}\int\frac{d^{2}q}{\left(2\pi\right)^{2}}\int_{0}^{\infty}\frac{d\varepsilon}{\pi}\left\{ \frac{\partial}{\partial\varepsilon}\left(\ln\left(1-e^{\beta\varepsilon}\right)-\beta\varepsilon\right)\right.\nonumber \\
 &  & \qquad\left.\times\mathrm{Im}\left(\ln\left[\beta\left(-\varepsilon+\xi\right)\right]\right)\right\} \nonumber \\
 & = & \frac{\partial}{\partial\beta}\int\frac{d^{2}q}{\left(2\pi\right)^{2}}\int_{0}^{\infty}\frac{d\varepsilon}{\pi}\left\{ \left(\ln\left(1-e^{\beta\varepsilon}\right)-\beta\varepsilon\right)\right.\nonumber \\
 &  & \qquad\left.\times\mathrm{Im}\left(\frac{1}{-\varepsilon+\xi}\right)\right\} \nonumber \\
 & = & -\int\frac{d^{2}q}{\left(2\pi\right)^{2}}\int_{0}^{\infty}\frac{d\varepsilon}{\pi}\frac{\varepsilon}{e^{\beta\varepsilon}-1}\mathrm{Im}G\left(\varepsilon,\mathbf{q}\right).\label{eq:Internal_energy_imag_Green}
\end{eqnarray}
Keeping in mind that the imaginary part of the Green's function determines
the spectral function $A\left(\varepsilon,\mathbf{q}\right)$ via
\begin{equation}
-\frac{1}{\pi}\mathrm{Im}G\left(\varepsilon,\mathbf{q}\right)=A\left(\varepsilon,\mathbf{q}\right),
\end{equation}
and the density of states $\nu\left(\varepsilon\right)$ is given
by
\begin{equation}
\nu\left(\varepsilon\right)=\int\frac{d^{2}q}{\left(2\pi\right)^{2}}A\left(\varepsilon,\mathbf{q}\right),
\end{equation}
formula (\ref{eq:Internal_energy_imag_Green}) can be interpreted
as an energy average over the Bose-Einstein distribution weightened
by the density of states:
\begin{equation}
E=\int_{0}^{\infty}d\varepsilon\,\frac{\varepsilon\nu\left(\varepsilon\right)}{e^{\beta\varepsilon}-1}.
\end{equation}

\end{appendix}

\bibliographystyle{thesis-bib}
\bibliography{References}

\begin{thebibliography}{100}

\bibitem{Wang2013}
L.~Wang, I.~Meric, P.~Y. Huang, Q.~Gao, Y.~Gao, H.~Tran, T.~Taniguchi,
  K.~Watanabe, L.~M. Campos, D.~A. Muller, J.~Guo, P.~Kim, J.~Hone, K.~L.
  Shepard, and C.~R. Dean, \textit{One-dimensional electrical contact to a
  two-dimensional material}, Science \textbf{342}, 614 (2013).

\bibitem{deJong1995}
M.~J.~M. de~Jong and L.~W. Molenkamp, \textit{Hydrodynamic electron flow in
  high-mobility wires}, Phys. Rev. B \textbf{51}, 13389 (1995).

\bibitem{Sulpizio2019}
J.~A. Sulpizio, L.~Ella, A.~Rozen, J.~Birkbeck, D.~J. Perello, D.~Dutta,
  M.~Ben-Shalom, T.~Taniguchi, K.~Watanabe, T.~Holder, R.~Queiroz, A.~Principi,
  A.~Stern, T.~Scaffidi, A.~K. Geim, and S.~Ilani, \textit{Visualizing
  poiseuille flow of hydrodynamic electrons}, Nature \textbf{576}, 75 (2019).

\bibitem{Gusev2020stokes}
G.~Gusev, A.~Jaroshevich, A.~Levin, Z.~Kvon, and A.~Bakarov, \textit{Stokes
  flow around an obstacle in viscous two-dimensional electron liquid},
  Scientific Reports \textbf{10}, 1 (2020).

\bibitem{Bandurin2016}
D.~A. Bandurin, I.~Torre, R.~K. Kumar, M.~B. Shalom, A.~Tomadin, A.~Principi,
  G.~H. Auton, E.~Khestanova, K.~S. Novoselov, I.~V. Grigorieva, L.~A.
  Ponomarenko, A.~K. Geim, and M.~Polini, \textit{Negative local resistance
  caused by viscous electron backflow in graphene}, Science \textbf{351}, 1055
  (2016).

\bibitem{KrishnaKumar2017}
R.~K. Kumar, D.~A. Bandurin, F.~M.~D. Pellegrino, Y.~Cao, A.~Principi, H.~Guo,
  G.~H. Auton, M.~B. Shalom, L.~A. Ponomarenko, G.~Falkovich, K.~Watanabe,
  T.~Taniguchi, I.~V. Grigorieva, L.~S. Levitov, M.~Polini, and A.~K. Geim,
  \textit{Superballistic flow of viscous electron fluid through graphene
  constrictions}, Nature Physics \textbf{13}, 1182 (2017).

\bibitem{Gurzhi1963}
R.~N. Gurzhi, \textit{Minimum of resistance in impurity-free conductors}, Zh.
  Eksp. Teor. Fiz. \textbf{44}, 771 (1963).

\bibitem{Gurzhi1968}
R.~Gurhzi, \textit{Hydrodynamic effects in solids at low temperature}, Sov.
  Phys. Usp. \textbf{11}, 255 (1968).

\bibitem{Bistritzer2011moire}
R.~Bistritzer and A.~H. MacDonald, \textit{Moir{\'e} bands in twisted
  double-layer graphene}, Proceedings of the National Academy of Sciences
  \textbf{108}, 12233 (2011).

\bibitem{Cao2018_TBG_superconductor}
Y.~Cao, V.~Fatemi, S.~Fang, K.~Watanabe, T.~Taniguchi, E.~Kaxiras, and
  P.~Jarillo-Herrero, \textit{Unconventional superconductivity in magic-angle
  graphene superlattices}, Nature \textbf{556}, 43 (2018).

\bibitem{Cao2020strangeTBG}
Y.~Cao, D.~Chowdhury, D.~Rodan-Legrain, O.~Rubies-Bigorda, K.~Watanabe,
  T.~Taniguchi, T.~Senthil, and P.~Jarillo-Herrero, \textit{Strange metal in
  magic-angle graphene with near planckian dissipation}, Physical Review
  Letters \textbf{124}, 076801 (2020).

\bibitem{Mackenzie2017}
A.~P. Mackenzie, \textit{The properties of ultrapure delafossite metals}, Rep.
  Prog. Phys. \textbf{80}, 32501 (2017).

\bibitem{Moll2016}
P.~J.~W. Moll, P.~Kushwaha, N.~Nandi, B.~Schmidt, and A.~P. Mackenzie,
  \textit{Evidence for hydrodynamic electron flow in $\mathrm{PdCoO}_2$},
  Science \textbf{351}, 1061 (2016).

\bibitem{nandi2018_magnetotransport_delafossites}
N.~Nandi, T.~Scaffidi, P.~Kushwaha, S.~Khim, M.~E. Barber, V.~Sunko,
  F.~Mazzola, P.~D. King, H.~Rosner, P.~J. Moll, M.~König, J.~E. Moore,
  S.~Hartnoll, and A.~P. Mackenzie, \textit{{Unconventional magneto-transport
  in ultrapure PdCoO$_2$ and PtCoO$_2$}}, npj Quantum Materials \textbf{3}, 1
  (2018).

\bibitem{nandi2019size_effects_delafossites}
N.~Nandi, T.~Scaffidi, S.~Khim, P.~Kushwaha, J.~Moore, and A.~MacKenzie,
  \textit{{Size restricted magnetotransport in the non-magnetic delafossite
  metals PdCoO$_2$ and PtCoO$_2$}}, APS \textbf{2019}, P40 (2019).

\bibitem{Andreev2011}
A.~V. Andreev, S.~A. Kivelson, and B.~Spivak, \textit{Hydrodynamic description
  of transport in strongly correlated electron systems}, Phys. Rev. Lett.
  \textbf{106}, 256804 (2011).

\bibitem{Briskot2015}
U.~Briskot, M.~Sch\"utt, I.~V. Gornyi, M.~Titov, B.~N. Narozhny, and A.~D.
  Mirlin, \textit{Collision-dominated nonlinear hydrodynamics in graphene},
  Phys. Rev. B \textbf{92}, 115426 (2015).

\bibitem{DAgosta2006Turbulence}
R.~D'Agosta and M.~Di~Ventra, \textit{Hydrodynamic approach to transport and
  turbulence in nanoscale conductors}, Journal of Physics: Condensed Matter
  \textbf{18}, 11059 (2006).

\bibitem{delacretaz2017hydroCDW}
L.~V. Delacr{\'e}taz, B.~Gout{\'e}raux, S.~A. Hartnoll, and A.~Karlsson,
  \textit{Theory of hydrodynamic transport in fluctuating electronic charge
  density wave states}, Physical Review B \textbf{96}, 195128 (2017).

\bibitem{Eguiluz1976hydrodynamicPlasmons}
A.~Eguiluz and J.~Quinn, \textit{Hydrodynamic model for surface plasmons in
  metals and degenerate semiconductors}, Physical Review B \textbf{14}, 1347
  (1976).

\bibitem{Forster}
D.~Forster, \textit{Hydrodynamic Fluctuations, Broken Symmetry, and Correlation
  Functions}, CRC Press (2018). ISBN 978-0367091323.

\bibitem{galitski2018dynamo}
V.~Galitski, M.~Kargarian, and S.~Syzranov, \textit{Dynamo effect and
  turbulence in hydrodynamic weyl metals}, Physical review letters
  \textbf{121}, 176603 (2018).

\bibitem{grozdanov2019holography}
S.~Grozdanov, A.~Lucas, and N.~Poovuttikul, \textit{Holography and
  hydrodynamics with weakly broken symmetries}, Physical Review D \textbf{99},
  086012 (2019).

\bibitem{Holder2019}
T.~Holder, R.~Queiroz, T.~Scaffidi, N.~Silberstein, A.~Rozen, J.~A. Sulpizio,
  L.~Ella, S.~Ilani, and A.~Stern, \textit{Ballistic and hydrodynamic
  magnetotransport in narrow channels}, Phys. Rev. B \textbf{100}, 245305
  (2019).

\bibitem{Hui2020}
V.~O. A.~Hui, S.~Lederer and E.-A. Kim, \textit{Quantum aspects of hydrodynamic
  transport from weak electron-impurity scattering}, Phys. Rev. B \textbf{101},
  121107 (2020).

\bibitem{Link2018Out}
J.~M. Link, B.~N. Narozhny, E.~I. Kiselev, and J.~Schmalian,
  \textit{Out-of-bounds hydrodynamics in anisotropic dirac fluids}, Phys. Rev.
  Lett. \textbf{120}, 196801 (2018).

\bibitem{Lucas2018}
A.~Lucas and K.~C. Fong, \textit{Hydrodynamics of electrons in graphene},
  Journal of Physics: Condensed Matter \textbf{30}, 53001 (2018).

\bibitem{Lucas20182D}
A.~Lucas and S.~D. Sarma, \textit{Electronic sound modes and plasmons in
  hydrodynamic two-dimensional metals}, Phys. Rev. B \textbf{97}, 115449
  (2018).

\bibitem{LucasH2017}
A.~Lucas and S.~A. Hartnoll, \textit{Resistivity bound for hydrodynamic bad
  metals}, PNAS \textbf{43}, 11344 (2017).

\bibitem{Moessner2018}
R.~Moessner, P.~Surówka, and P.Witkowski, \textit{Pulsating flow and boundary
  layers in viscous electronic hydrodynamics}, Phys. Rev. B \textbf{97}, 161112
  (2018).

\bibitem{Narozhny2017}
B.~N. Narozhny, I.~V. Gornyi, and A.~D. Mirlin, \textit{Hydrodynamic approach
  to electronic transport in graphene}, Ann. d. Phys. \textbf{529}, 1700043
  (2017).

\bibitem{Principi2015}
A.~Principi and G.~Vignale, \textit{Violation of the wiedemann-franz law in
  hydrodynamic electron liquids}, Phys. Rev. Lett. \textbf{115}, 056603 (2015).

\bibitem{Scaffidi2017}
T.~Scaffidi, N.~Nandi, B.~Schmidt, A.~P. Mackenzie, and J.~E. Moore,
  \textit{Hydrodynamic electron flow and hall viscosity}, Phys. Rev. Lett.
  \textbf{118}, 226601 (2017).

\bibitem{Svintsov2018}
D.~Svintsov, \textit{Hydrodynamic-to-ballistic crossover in dirac materials},
  Phys. Rev. B \textbf{97}, 121405 (2018).

\bibitem{Svintsov2019}
D.~Svintsov, \textit{Emission of plasmons by drifting dirac electrons: A
  hallmark of hydrodynamic transport}, Phys. Rev. B \textbf{100}, 195428
  (2019).

\bibitem{Varnavides2020anisotropic_electron_hydro}
G.~Varnavides, A.~S. Jermyn, P.~Anikeeva, C.~Felser, and P.~Narang,
  \textit{Electron hydrodynamics in anisotropic materials}, Nature
  Communications \textbf{11}, 1 (2020).

\bibitem{Zdyrski2019}
T.~Zdyrski and J.~McGreevy, \textit{Effects of dissipation on solitons in the
  hydrodynamic regime of graphene}, Phys. Rev. B \textbf{99}, 235435 (2019).

\bibitem{lucas2015memory}
A.~Lucas and S.~Sachdev, \textit{Memory matrix theory of magnetotransport in
  strange metals}, Physical Review B \textbf{91}, 195122 (2015).

\bibitem{buchel2019_holographic_hydrodynamic}
A.~Buchel and M.~Baggioli, \textit{Holographic viscoelastic hydrodynamics},
  Journal of High Energy Physics \textbf{2019}, 146 (2019).

\bibitem{Dyakonov_Furman_1987charge_relaxation}
M.~D'yakonov and A.~Furman, \textit{Charge relaxation in an anisotropic medium
  and in low-dimensional media}, Zh. Eksp. Teor. Fiz \textbf{92}, 1012 (1987).

\bibitem{Liu2007_Yukawa_superdiffusion}
B.~Liu and J.~Goree, \textit{Superdiffusion in two-dimensional yukawa liquids},
  Physical Review E \textbf{75}, 016405 (2007).

\bibitem{Ott2009_Diffusion}
T.~Ott and M.~Bonitz, \textit{Is diffusion anomalous in two-dimensional yukawa
  liquids?}, Physical review letters \textbf{103}, 195001 (2009).

\bibitem{Note1}
See this Ref. for a review of literature on anomalous diffusion in dusty
  plasmas.

\bibitem{Ott2014_One_component_Coulomb}
T.~Ott, M.~Bonitz, L.~Stanton, and M.~Murillo, \textit{Coupling strength in
  coulomb and yukawa one-component plasmas}, Physics of Plasmas \textbf{21},
  113704 (2014).

\bibitem{Warren2000_Colloidal_Suspensions}
P.~B. Warren, \textit{A theory of void formation in charge-stabilized colloidal
  suspensions at low ionic strength}, The Journal of Chemical Physics
  \textbf{112}, 4683 (2000).

\bibitem{kwasnicki2017_Fractional_laplace}
M.~Kwa{\'s}nicki, \textit{{Ten equivalent definitions of the fractional Laplace
  operator}}, Fractional Calculus and Applied Analysis \textbf{20}, 7 (2017).

\bibitem{Samko1993}
S.~G. Samko, A.~A. Kilbas, and O.~I. Marichev, \textit{Fractional integrals and
  derivatives}, Taylor \& Francis (1993). ISBN: 978-2881248641.

\bibitem{Bouchaud1990}
J.-P. Bouchaud and A.~Georges, \textit{Anomalous diffusion in disordered media:
  Statistical mechanisms, models and physical applications}, Physics Reports
  \textbf{195}, 127 (1990).

\bibitem{Gnedenko1954}
B.~V. Gnedenko and A.~N. Kolmogorov, \textit{Limit Distributions for Sums of
  Independent Random Variables}, Adison Wesley (1954).

\bibitem{Metzler1999}
R.~Metzler, E.~Barkai, and J.~Klafter, \textit{Anomalous diffusion and
  relaxation close to thermal equilibrium: A fractional fokker-planck equation
  approach}, Phys. Rev. Lett. \textbf{82}, 3563 (1999).

\bibitem{Metzler2000}
R.~Metzler and J.~Klafter, \textit{The random walk's guide to anomalous
  diffusion: a fractional dynamics approach}, Phys. Rep. \textbf{339}, 1
  (2000).

\bibitem{Lucas2016sound_modes_neutr_graphene}
A.~Lucas, \textit{Sound waves and resonances in electron-hole plasma}, Physical
  Review B \textbf{93}, 245153 (2016).

\bibitem{kolomeisky2017relaxation}
E.~B. Kolomeisky and J.~P. Straley, \textit{Relaxation of charge in monolayer
  graphene: Fast nonlinear diffusion versus coulomb effects}, Physical Review B
  \textbf{95}, 045415 (2017).

\bibitem{Volkov2016magnetoplasmon_retardation_effects}
V.~Volkov and A.~Zabolotnykh, \textit{Undamped relativistic magnetoplasmons in
  lossy two-dimensional electron systems}, Physical Review B \textbf{94},
  165408 (2016).

\bibitem{Hoepfel1982_Thermal_plasmons}
R.~A. H{\"o}pfel, E.~Vass, and E.~Gornik, \textit{Thermal excitation of
  two-dimensional plasma oscillations}, Physical Review Letters \textbf{49},
  1667 (1982).

\bibitem{Sun2018_SpecificHeat_Sr3Ru2O7}
D.~Sun, A.~Rost, R.~Perry, A.~Mackenzie, and M.~Brando, \textit{Low temperature
  thermodynamic investigation of the phase diagram of {$Sr_{3}Ru_{2}O_{7}$}},
  Physical Review B \textbf{97}, 115101 (2018).

\bibitem{Millis1993_Temperature_quantum_critical}
A.~Millis, \textit{Effect of a nonzero temperature on quantum critical points
  in itinerant fermion systems}, Physical Review B \textbf{48}, 7183 (1993).

\bibitem{Mousatov2020_strange_metal_Sr3Ru2O7}
C.~H. Mousatov, E.~Berg, and S.~A. Hartnoll, \textit{Theory of the strange
  metal {$Sr_{3}Ru_{2}O_{7}$}}, Proceedings of the National Academy of Sciences
  \textbf{117}, 2852 (2020).

\bibitem{Mittendorff2014pump_probe_anisotropy}
M.~Mittendorff, T.~Winzer, E.~Malic, A.~Knorr, C.~Berger, W.~A. de~Heer,
  H.~Schneider, M.~Helm, and S.~Winnerl, \textit{Anisotropy of excitation and
  relaxation of photogenerated charge carriers in graphene}, Nano letters
  \textbf{14}, 1504 (2014).

\bibitem{houston1977time_of_flight}
P.~Houston and A.~Evans, \textit{Electron drift velocity in n-gaas at high
  electric fields}, Solid-State Electronics \textbf{20}, 197 (1977).

\bibitem{hopfel1986picosecond_time_of_flight}
R.~H{\"o}pfel, J.~Shah, D.~Block, and A.~Gossard, \textit{Picosecond
  time-of-flight measurements of minority electrons in gaas/algaas quantum well
  structures}, Applied physics letters \textbf{48}, 148 (1986).

\bibitem{LLHydro}
L.~D. Landau and E.~M. Lifshits, \textit{Fluid mechanics},
  Butterworth-Heinemann (1987). ISBN 978-0750627672.

\bibitem{Batchelor2000}
G.~Batchelor, \textit{An introduction to fluid dynamics}, Cambridge University
  Press (2000). ISBN 978-0521663960.

\bibitem{silin1958_landau_silin_theory}
V.~Silin, \textit{Theory of a degenerate electron liquid}, JETP \textbf{6}, 387
  (1958).

\bibitem{Vlasov1938}
A.~A. Vlasov, \textit{On vibration properties of electron gas}, J. Exp. Theor.
  Phys. \textbf{8}, 291 (1938).

\bibitem{stern1967polarizability}
F.~Stern, \textit{Polarizability of a two-dimensional electron gas}, Physical
  Review Letters \textbf{18}, 546 (1967).

\bibitem{zala2001interaction}
G.~Zala, B.~Narozhny, and I.~Aleiner, \textit{Interaction corrections at
  intermediate temperatures: Longitudinal conductivity and kinetic equation},
  Physical Review B \textbf{64}, 214204 (2001).

\bibitem{giona1992fractional_random_media}
M.~Giona and H.~E. Roman, \textit{Fractional diffusion equation for transport
  phenomena in random media}, Physica A: Statistical Mechanics and its
  Applications \textbf{185}, 87 (1992).

\bibitem{scalas2000fractional_finance}
E.~Scalas, R.~Gorenflo, and F.~Mainardi, \textit{Fractional calculus and
  continuous-time finance}, Physica A: Statistical Mechanics and its
  Applications \textbf{284}, 376 (2000).

\bibitem{Baggioli2020Relativistic_Diffusion_Fractional}
M.~Baggioli, G.~La~Nave, and P.~W. Phillips, \textit{Fractional hydrodynamics
  and anomalous diffusion}, arXiv preprint arXiv:2006.10064  (2020).

\bibitem{Desbois1992_2d_Levy}
J.~Desbois, \textit{{Algebraic areas distributions for two-dimensional Levy
  flights}}, Journal of Physics A: Mathematical and General \textbf{25}, L755
  (1992).

\bibitem{Kiselev2020}
E.~I. Kiselev and J.~Schmalian, \textit{Nonlocal hydrodynamic transport and
  collective excitations in dirac fluids}, Physical Review B \textbf{102},
  245434 (2020).

\bibitem{Fritz2008}
L.~Fritz, J.~Schmalian, M.~Müller, and S.~Sachdev, \textit{Quantum critical
  transport in clean graphene}, Phys. Rev. B \textbf{78}, 85416 (2008).

\bibitem{Kiselev2019b}
E.~I. Kiselev and J.~Schmalian, \textit{L\'evy flights and hydrodynamic
  superdiffusion on the dirac cone of graphene}, Phys. Rev. Lett. \textbf{123},
  195302 (2019).

\bibitem{Pizarro2019TBG_screening}
J.~Pizarro, M.~R{\"o}sner, R.~Thomale, R.~Valent{\'\i}, and T.~Wehling,
  \textit{Internal screening and dielectric engineering in magic-angle twisted
  bilayer graphene}, Physical Review B \textbf{100}, 161102(R) (2019).

\bibitem{Goodwin2019TBG_screening_attractive}
Z.~A. Goodwin, F.~Corsetti, A.~A. Mostofi, and J.~Lischner, \textit{Attractive
  electron-electron interactions from internal screening in magic-angle twisted
  bilayer graphene}, Physical Review B \textbf{100}, 235424 (2019).

\bibitem{hamaguchi1994_thermodynamics_Yukawa}
S.~Hamaguchi and R.~Farouki, \textit{Thermodynamics of strongly-coupled yukawa
  systems near the one-component-plasma limit. i. derivation of the excess
  energy}, The Journal of chemical physics \textbf{101}, 9876 (1994).

\bibitem{Feng2013_Dusty_plasma_visc}
Y.~Feng, J.~Goree, and B.~Liu, \textit{Longitudinal viscosity of
  two-dimensional yukawa liquids}, Physical Review E \textbf{87}, 013106
  (2013).

\bibitem{kalman2004_2D_Yukawa}
G.~Kalman, P.~Hartmann, Z.~Donk{\'o}, and M.~Rosenberg, \textit{Two-dimensional
  yukawa liquids: Correlation and dynamics}, Physical review letters
  \textbf{92}, 065001 (2004).

\bibitem{PinesNozieres1}
D.~Pines and P.~Nozi\`eres, \textit{Theory Of Quantum Liquids: Normal Fermi
  Liquids}, CRC Press (1989). ISBN 978-0201407747.

\bibitem{efros2008_Einstein_Relation}
A.~Efros, \textit{Negative density of states: Screening, einstein relation, and
  negative diffusion}, Physical Review B \textbf{78}, 155130 (2008).

\bibitem{Shur2013_GaAs_Devices}
M.~S. Shur, \textit{GaAs devices and circuits}, Springer Science \& Business
  Media (2013).

\bibitem{Nguyen2019Gating}
P.~V. Nguyen, N.~C. Teutsch, N.~P. Wilson, J.~Kahn, X.~Xia, A.~J. Graham,
  V.~Kandyba, A.~Giampietri, A.~Barinov, G.~C. Constantinescu \textit{et~al.},
  \textit{Visualizing electrostatic gating effects in two-dimensional
  heterostructures}, Nature \textbf{572}, 220 (2019).

\bibitem{Dyakonov1993_Dyakonov_Shur_Instability_gated}
M.~Dyakonov and M.~Shur, \textit{Shallow water analogy for a ballistic field
  effect transistor: New mechanism of plasma wave generation by dc current},
  Physical review letters \textbf{71}, 2465 (1993).

\bibitem{Dyakonov1996_Nonlinear_Mixing}
M.~Dyakonov and M.~Shur, \textit{Detection, mixing, and frequency
  multiplication of terahertz radiation by two-dimensional electronic fluid},
  IEEE transactions on electron devices \textbf{43}, 380 (1996).

\bibitem{Torre2019}
I.~Torre, L.~V. de~Castro, B.~V. Duppen, D.~B. Ruiz, F.~M. Peeters, F.~H.~L.
  Koppens, and M.~Polini, \textit{Acoustic plasmons at the crossover between
  the collisionless and hydrodynamic regimes in two-dimensional electron
  liquids}, Phys. Rev. B \textbf{99}, 144307 (2019).

\bibitem{Zabolotnykh2019_GatedPlasmons}
A.~Zabolotnykh and V.~Volkov, \textit{Interaction of gated and ungated plasmons
  in two-dimensional electron systems}, Physical Review B \textbf{99}, 165304
  (2019).

\bibitem{Horing1976_Magnetoplasmon}
N.~J.~M. Horing and M.~M. Yildiz, \textit{Quantum theory of longitudinal
  dielectric response properties of a two-dimensional plasma in a magnetic
  field}, Annals of Physics \textbf{97}, 216 (1976).

\bibitem{Mast1985_Magnetoplasmon}
D.~Mast, A.~Dahm, and A.~Fetter, \textit{Observation of bulk and edge
  magnetoplasmons in a two-dimensional electron fluid}, Physical review letters
  \textbf{54}, 1706 (1985).

\bibitem{Chandrasekhar1960}
S.~Chandrasekhar, \textit{Stochastic problems in physics and astronomy}, Rev.
  Mod. Phys. \textbf{15}, 1 (1943).

\bibitem{Uuguz2009_Colloidal_Suspensions}
E.~C. O{\u{g}}uz, R.~Messina, and H.~L{\"o}wen, \textit{Multilayered crystals
  of macroions under slit confinement}, Journal of Physics: Condensed Matter
  \textbf{21}, 424110 (2009).

\bibitem{Trachenko2015collective_thermodynamics}
K.~Trachenko and V.~Brazhkin, \textit{Collective modes and thermodynamics of
  the liquid state}, Reports on Progress in Physics \textbf{79}, 016502 (2015).

\bibitem{Baggioli2019BosonPeak_diffusive}
M.~Baggioli and A.~Zaccone, \textit{Universal origin of boson peak vibrational
  anomalies in ordered crystals and in amorphous materials}, Physical review
  letters \textbf{122}, 145501 (2019).

\bibitem{Baggioli2019LinearTCv_diffusion}
M.~Baggioli and A.~Zaccone, \textit{Hydrodynamics of disordered marginally
  stable matter}, Physical Review Research \textbf{1}, 012010 (2019).

\bibitem{Belitz1997Nonanalytic_FL}
D.~Belitz, T.~R. Kirkpatrick, and T.~Vojta, \textit{Nonanalytic behavior of the
  spin susceptibility in clean fermi systems}, Physical Review B \textbf{55},
  9452 (1997).

\bibitem{Chubukov2003Nonanalytic_FL}
A.~V. Chubukov and D.~L. Maslov, \textit{Nonanalytic corrections to the
  fermi-liquid behavior}, Physical Review B \textbf{68}, 155113 (2003).

\bibitem{Chubukov2004Singular_FL}
A.~V. Chubukov and D.~L. Maslov, \textit{Singular corrections to the
  fermi-liquid theory}, Physical Review B \textbf{69}, 121102 (2004).

\bibitem{Chubukov2005_Singular_2D}
A.~V. Chubukov, D.~L. Maslov, S.~Gangadharaiah, and L.~I. Glazman,
  \textit{Singular perturbation theory for interacting fermions in two
  dimensions}, Physical Review B \textbf{71}, 205112 (2005).

\bibitem{AltlandSimons}
A.~Altland and B.~D. Simons, \textit{Condensed matter field theory}, Cambridge
  university press (2010). ISBN: 978-0521769754.

\bibitem{negele2018quantum}
J.~W. Negele and H.~Orland, \textit{Quantum many-particle systems}, CRC Press
  (2018).

\end{thebibliography}

\end{document}